\documentclass[aps,prl,superscriptaddress,twocolumn,longbibliography,floatfix]{revtex4-2}
\usepackage{units}
\usepackage{amsmath}
\usepackage{amsthm}
\usepackage{amssymb}
\usepackage{graphicx}
\usepackage{color}
\usepackage{xcolor}
\usepackage{bbold}
\usepackage{titlesec} 
\usepackage{mathrsfs}
\usepackage{braket}
\usepackage{float}

\definecolor{myurlcolor}{rgb}{0,0,0.7}
\definecolor{myrefcolor}{rgb}{0.1,0,0.9}

\usepackage[
	breaklinks,
	pdftex,
	colorlinks=true, 
	linkcolor=myrefcolor,
	citecolor=myurlcolor,
	urlcolor=myurlcolor
]{hyperref}

\usepackage[linesnumbered, ruled,vlined]{algorithm2e}

\graphicspath{{./images/},{./imagesAppendix/}}

\def\app#1#2{%
  \mathrel{%
    \setbox0=\hbox{$#1\sim$}%
    \setbox2=\hbox{%
      \rlap{\hbox{$#1\propto$}}%
      \lower1.1\ht0\box0%
    }%
    \raise0.25\ht2\box2%
  }%
}

\ifx\proof\undefined
\newenvironment{proof}[1][\protect\proofname]{\par
	\normalfont\topsep6\p@\@plus6\p@\relax
	\trivlist
	\itemindent\parindent
	\item[\hskip\labelsep\scshape #1]\ignorespaces
}{%
	\endtrivlist\@endpefalse
}
\providecommand{\proofname}{Proof}
\fi
\makeatother

\newcommand{\tr}{\mathrm{tr}}

\providecommand{\factname}{Fact}
\providecommand{\theoremname}{Theorem}
\providecommand{\claimname}{Claim}
\providecommand{\lemmaname}{Lemma}
\providecommand{\definitionname}{Definition}
\providecommand{\corollaryname}{Corollary}

\definecolor{KB}{rgb}{0.4,0.3,0.9}

\definecolor{THc}{rgb}{0.9,0.3,0.2}

\newcommand{\be}{\begin{equation}}
\newcommand{\ee}{\end{equation}}
\newcommand{\ba}{\begin{eqnarray}}
\newcommand{\ea}{\end{eqnarray}}

\newcommand{\poly}{\operatorname{poly}}

\newcommand{\st}[1]{\ket{#1}\!\!\bra{#1}}

\newcommand{\sectionMain}[1]{
\let\oldaddcontentsline\addcontentsline
\renewcommand{\addcontentsline}[3]{}
\section{#1}
\let\addcontentsline\oldaddcontentsline
}

\newlength{\fighskip} \fighskip=2pt
\newlength{\figvskip} \figvskip=1pt

\makeatletter
\def\namedlabel#1#2{\begingroup
   \def\@currentlabel{#2}%
   \label{#1}\endgroup
}
\makeatother
\newcommand{\prlsection}[1]{{\em {#1}.---}}

\allowdisplaybreaks

\begin{document}

\title{Unscrambling Quantum Information with Clifford Decoders}

\author{Salvatore F.E. Oliviero}\email{s.oliviero001@umb.edu}
\affiliation{Physics Department,  University of Massachusetts Boston, Massachusetts 02125, USA}
\affiliation{Theoretical Division (T-4), Los Alamos National Laboratory, Los Alamos, New Mexico 87545, USA}
\affiliation{Center for Nonlinear Studies, Los Alamos National Laboratory, Los Alamos, New Mexico 87545, USA}

\author{Lorenzo Leone}\email{lorenzo.leone001@umb.edu}
\affiliation{Physics Department,  University of Massachusetts Boston,Massachusetts  02125, USA}
\affiliation{Theoretical Division (T-4), Los Alamos National Laboratory, Los Alamos, New Mexico 87545, USA}
\affiliation{Center for Nonlinear Studies, Los Alamos National Laboratory, Los Alamos, New Mexico 87545, USA}

\author{{Seth} {Lloyd}}\email{slloyd@mit.edu}
\affiliation{Department of Mechanical Engineering, Massachusetts Institute of Technology,  {Cambridge},  {Massachusetts}, {USA}}
\affiliation{Turing Inc., Brooklyn, New York, USA}

\author{Alioscia Hamma}\email{alioscia.hamma@unina.it}

\affiliation{Dipartimento di Fisica `Ettore Pancini', Universit\`a degli Studi di Napoli Federico II,
Via Cintia 80126,  Napoli, Italy}
\affiliation{INFN, Sezione di Napoli, Italy}

\begin{abstract}

Quantum information scrambling is a unitary process that destroys local correlations and spreads information throughout the system, effectively hiding it in nonlocal degrees of freedom. In principle, unscrambling this information is possible with perfect knowledge of the unitary dynamics~[B. Yoshida and A.Kitaev, \href{https://arxiv.org/abs/1710.03363}{arXiv:1710.03363}.]. However, this Letter demonstrates that even without previous knowledge of the internal dynamics, information can be efficiently decoded from an unknown scrambler by monitoring the outgoing information of a local subsystem. We show that rapidly mixing but not fully chaotic scramblers can be decoded using Clifford decoders. The essential properties of a scrambling unitary can be efficiently recovered, even if the process is exponentially complex. Specifically, we establish that a unitary operator composed of $t$ non-Clifford gates admits a Clifford decoder up to $t\le n$.

\end{abstract}

\maketitle

\prlsection{Introduction}%
 ~A famous English nursery rhyme~\footnote{Humpty Dumpty sat on a wall, Humpty Dumpty had a great fall. All the king's horses and all the king's men Couldn't put Humpty together again.}
states that, once an egg is broken, it is quite arduous to put it together, no matter how many resources the king may employ. At the quantum mechanical level, breaking an egg corresponds to information scrambling~\cite{hosur2016ChaosQuantumChannels,ding2016ConditionalMutualInformation,brown2013ScramblingSpeedRandom,liu2018GeneralizedEntanglementEntropies,liu2018EntanglementQuantumRandomness,styliaris2021InformationScramblingBipartitions}, that is, the spreading of information---initially localized in a part of a quantum system---into quantum correlations all over the entire system.

The primary consequence of quantum information scrambling is that no local measurements can fully reconstruct the scrambled information. This phenomenon has been extensively explored, e.g. in the context of black hole physics~\cite{lloyd1988BlackHolesDemons,page1993InformationBlackHole,hayden2007BlackHolesMirrors,shenker2014BlackHolesButterfly,shenker2015StringyEffectsScrambling, roberts2015LocalizedShocks, shenker2014MultipleShocks, shenker2015StringyEffectsScrambling}: after information falls into a black hole, it cannot be recovered solely by examining the outgoing Hawking radiation. In this context, it has been conjectured that black holes are fast scramblers~\cite{sekino2008FastScramblers,lashkari2013FastScramblingConjecture,maldacena2016BoundChaos}, i.e., within a time $\tau^{*}=O(\log n)$, which scales logarithmically with the number $n$ of systems' degrees of freedom, the information spreads nonlocally.

The scrambling capability of a unitary dynamics  can be probed by the decay of out-of-time-order correlators (OTOCs)~\cite{kitaev2014HiddenCorrelationsHawking} that capture the sensitivity of the dynamics to local perturbations and is, as such, the quantum equivalent of the butterfly effect\cite{roberts2016LiebRobinsonBoundButterfly}. A scrambler $U_t$ can be realized by a Clifford circuit doped by a number $t$ of non-Clifford resources~\cite{zhou2020SingleGateClifford,leone2021QuantumChaosQuantum,oliviero2021TransitionsEntanglementComplexity,true2022TransitionsEntanglementComplexity}. Clifford unitaries $U_0$ are structurally very simple; indeed,  they can be both represented and learned by polynomial resources~\cite{low2009LearningTestingAlgorithms,lai2022LearningQuantumCircuits}. In spite of this, they can be 
fast  scramblers~\cite{chamon2022QuantumStatisticalMechanics, roberts2017ChaosComplexityDesign,oliviero2021RandomMatrixTheory,leone2021IsospectralTwirlingQuantum}.
On the other hand, doped unitaries $U_t$  become exponentially more complex in $t$ to be represented and simulated~\cite{aaronson2004ImprovedSimulationStabilizer,bravyi2016ImprovedClassicalSimulation}.  

Although scrambling destroys local correlations, in a seminal paper~\cite{hayden2007BlackHolesMirrors}, it was shown that local quantum information tossed in the input of a scrambler can actually be recovered by measuring a local output subsystem, of size slightly larger compared to the scrambled information. However, this successful recovery relies on the precise knowledge of $U^{\dag}_t$~\cite{yoshida2017EfficientDecodingHaydenPreskill}, which allows for the construction of a decoder capable of distilling back the scrambled information, effectively reversing the scrambling process.



In this Letter, we relax the assumption of perfect knowledge of the scrambler dynamics $U_t$, which can be too strong in many contexts of interest, and pose the question of whether one can \textit{learn}, after tossing in known \textit{test} states, how to retrieve the scrambled information by solely observing a local output subsystem, without any previous knowledge of the dynamics $U_t$. 

Informally, the main result of this work is to show that this is indeed possible. The learning cost is exponential in $t$, but, and this is striking,  the decoder $V$ is in itself a simple Clifford operator, even for a very complex $U_t$. It turns out that the Clifford part of $U_t$ is sufficient to decode the information of $U_t$. This can be efficiently encoded in $V$, while all its complexity (given by the injection of non-Clifford resources) turns out to be useless.

More precisely, the algorithm presented in this Letter learns a Clifford decoder $V$ for $U_t$ by $\poly(n,2^t)$ query accesses to $U_t$~\footnote{we refer to ``query access'' as the ability to perform the unitary transformation $U_t$ followed by a measurement on a quantum register consisting of $n$ qubits}.  
The fidelity $\mathcal{F}(V)$ of the retrieved information by $V$ is 
\begin{equation}\label{mainF}
\mathcal{F}(V)\ge \frac{1}{1+2^{2|A|+t-2|D|}}\,,
\end{equation}
while the probability $\mathscr{P}(V)$ of learning  the decoder $V$ is 
\begin{eqnarray}
\label{mainP}
\mathscr{P}(V)\ge 1-2^{t-2(n-|D|)}\,.
\label{successprobability}
\end{eqnarray}
\noindent
Here $n$ is the total number of qubits of the scrambler $U_t$, $|A|$ is the number of qubits of input information, $|D|$ is the number of readout qubits.


Equations.~\eqref{mainF} and~\eqref{successprobability} also set the domain of effectiveness of the algorithm. 
As long as $t\le n$, which we refer to as ``quasichaotic regime''~\cite{leone2022LearningEfficientDecoders}, the unitary $U_t$ is exponentially complex, yet one can still learn its efficient Clifford decoder at the cost of linearly increasing the size of the readout qubits $D$. However, approaching the quasichaotic regime, the number of readout qubits increases with $t$, indicating a more complex and mixing scrambling process that seems to be not locally reversed. As soon as $t>n$, we observe that both the fidelity~\eqref{mainF} and the probability of learning~\eqref{successprobability} decay exponentially in $t$ for any choice of $|D|$, until one reaches full quantum chaos at $t\ge 2n$~\cite{leone2021QuantumChaosQuantum}, where the unitary $U_t$ resembles the properties of a random unitary operator, and for which the fidelity and learning probability are exponentially small in $n$.

\begin{figure*}[t]
\includegraphics[width=\textwidth]{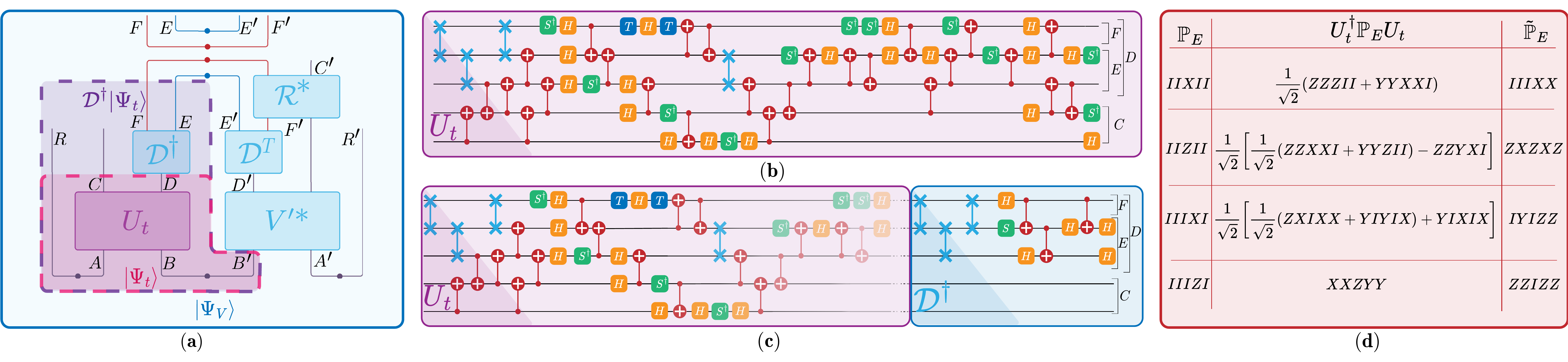}
\caption{\raggedright{ (\textbf a) diagrammatic representation of $\ket{\Psi_V}$, where the upward direction represents the progression of time. The steps of the decoding algorithm generating $\ket{\Psi}_V$ are:  (1) parallel application of $U_t$ (obtaining so $\ket{\Psi_t}$, diagrammatically shown in the purple box) and of the decrypter $V^\prime$ on the initial state $\ket{RA}\ket{BB^\prime}\ket{A^\prime R^\prime}$ (2) application of  $\mathcal{D}^{\dag}$ (obtaining $\mathcal{D}\ket{\Psi}_t$, as diagrammatically shown in the violet box) and $\mathcal{D}^{T}$. This process dumps the stabilizer entropy onto the $F, F^\prime$ subspaces. (3) The final steps involve the application of $\mathcal R^*$ and of a projective measurement on $DD^\prime$. Panel (\textbf{b}) provides an example of a doped random Clifford circuit $U_t$, whereas panel (\textbf{c}) displays the circuit $\mathcal{D}^{\dagger}U_t$, where $\mathcal{D}$ is a diagonalizer for the circuit $U_t$ given in panel (\textbf b). Panel (\textbf{d}) illustrates the adjoint action of both circuits on the generators of the group $\mathbb{P}_E$. The circuit $U_t$ shown in panel (\textbf{b}) only preserves the generator $IIZII$, transforming it into another Pauli operator, whereas the adjoint action of $\mathcal{D}^\dagger U_t$ preserves all generators, showing how the diagonalizer can move non-Cliffordness away from the subsystem of interest.  }}
\label{fig:diagram}
\end{figure*}

\prlsection{Decoding scramblers}  In this Letter, a scrambler is a unitary $U_t$ acting on a joint system $A\cup B$ of $n=|A|+|B|$ qubits with output  $C\cup D$, i.e., $U_t:A\cup B\mapsto C\cup D$ (see Fig.~\ref{fig:diagram}). The initially localized information is represented by the state in subsystem $A$ while subsystem $D$ represents the readable output subsystem. 

Let us formally define a scrambling unitary. Consider two subsystems $X$ and $Y$, the OTOC $\Omega_{XY}(U_t)$ is defined as
\be
\Omega_{XY}(U_t)=\frac{1}{2^n}\langle \tr(P_X U_t^{\dag}P_YU_tP_X U_t^{\dag}P_YU_t)\rangle_{X,Y}\,,\label{otocdef}
\ee
where $P_X, P_Y$ are Pauli strings with support on $X$ and $Y$, respectively, and $\braket{\cdot}_{X,Y}$ represents the average with respect to the local Pauli groups, i.e., the local observables, on $X$ and $Y$.
The unitary $U_t$ is a scrambler if and only if the OTOCs $\Omega_{XY}(U_t)$ behave as~\cite{yoshida2017EfficientDecodingHaydenPreskill}:
\be\label{scramblingdef}
\Omega_{XY}(U_t)\simeq \frac{1}{2^{2|X|}}+\frac{1}{2^{2|Y|}}-\frac{1}{2^{2(|X|+|Y|)}}\,.
\ee



Let us now describe an adversarial setup often used in the context of information scrambling~\cite{hayden2007BlackHolesMirrors,yoshida2017EfficientDecodingHaydenPreskill,yoshida2019DisentanglingScramblingDecoherence}. We consider Alice $R$ and Bob $B^{\prime}$ sharing respectively an EPR pair (Bell pair) with the input state $AB$ of the scrambler $U_t$, i.e., the state of the whole system $ARBB^{\prime}$ is \be
\ket{\Psi_t}\equiv U_{t}\ket{BB^{\prime}}\ket{AR}\,,
\label{statescrambled}
\ee 
where we denote $\ket{\Lambda\Lambda^{\prime}}=2^{-|\Lambda|/2}\sum_{i=1}^{2^{|\Lambda|}}\ket{i}_{\Lambda}\ket{i}_{\Lambda^{\prime}}$ an EPR pair between $\Lambda$ and $\Lambda^{\prime}$ and $\Pi_{\Lambda\Lambda^{\prime}}\equiv \st{\Lambda\Lambda^{\prime}}$.

One then questions how much (local) correlation between $A$ and $R$ survives after the unitary dynamics $U_t$. In this regard, the decoupling theorem~\cite{hayden2007BlackHolesMirrors} states that, after the scrambling dynamics $U_t$, 
the mutual information $I(R|DB^{\prime})\equiv |A|+\log 2^{2|A|}\Omega_{AD}(U_t)$ between $R$ and $D\cup B^{\prime}$ for \textit{any }$D$ such that $|D|=|A|+\log \epsilon^{-1/2}$ is, thanks to Eq.~\eqref{scramblingdef}, $\epsilon$-maximal 
\begin{equation}
I(R|DB^{\prime})=|A|-\epsilon\,,\label{decouplingth}
\end{equation}
and thus $A$ is completely \textit{decoupled} from $R$, i.e., $I(R|A)\equiv|A|-I(R|DB^{\prime})=\epsilon$. Since now the information is perfectly correlated with Bob's qubits, there exists a unitary $V$ that \textit{decodes} the information and enables Bob to distill an EPR pair between Alice $R$ and a reference system of the same dimension of $R$, say $R^{\prime}$. As a result, Bob can access all the information in Alice's possession by just looking at $B^{\prime}$ together with \textit{any} subsystem $D$ containing slightly more qubits than the ones in $A$. 

As shown in~\cite{yoshida2017EfficientDecodingHaydenPreskill}, to read the output subsystem $D$, Bob first appends a reference state $\ket{A^{\prime}R^{\prime}}$, applies the decoder $V\,:\,A^{\prime}\cup B^{\prime}\mapsto C^{\prime}\cup D^{\prime}$, and then projects the resulting state onto $\ket{DD^{\prime}}$. After decoding, we obtain the final state
$\ket{\Psi_V}={{\pi_V}}^{-1/2}\Pi_{DD^{\prime}}V^{T}\ket{\Psi_t}\ket{A^{\prime}R^{\prime}}$, where $\pi_V$ is a normalization and $T$ is the transposition. The fidelity $\mathcal{F}(V)$ between $\ket{\Psi_V}$ and the target EPR pair $\ket{RR^{\prime}}$ quantifies the success of the decoding protocol by Bob and is defined as
$
\mathcal{F}(V)\equiv\bra{\Psi_V}\Pi_{RR^{\prime}}\ket{\Psi_{V}}
$. The closer $\mathcal{F}(V)$ is to one, the better the decoding.

In~\cite{yoshida2017EfficientDecodingHaydenPreskill} it has been proven that a perfect decoder $V$ would be the inverse scrambling operation $U^{\dag}_{t}$, i.e., $\mathcal{F}(U_{t}^{\dag})=1-\epsilon$. However, as we will see, a perfect decoder is not unique. There are potentially an infinite number of perfect decoders $V$ with possibly very low gate fidelity with $U_t$.


\prlsection{The Clifford decoder}
Let us show in a simplified fashion how the perfect decoding of a scrambler $U_t$ can be achieved by a Clifford decoder $V$. We will see that the Clifford decoder $V$ can be written as
\be
V=\mathcal{D}\mathcal{R}\mathcal{D}^{\dag}V^{\prime}\label{thedecoder}
\ee
i.e., as a product of three Clifford operators, namely the ``diagonalizer'' $\mathcal{D}$, the ``randomizer'' $\mathcal{R}$, and the ``decrypter'' $V^{\prime}$. While here we operate under the assumption that the knowledge of $V$ is given and explore the role of each component $\mathcal{D},\mathcal{R},V^{\prime}$ in Eq.~\eqref{thedecoder}. In the subsequent section, we will elaborate on how Bob can learn the Clifford decoder $V$ from the output of the scrambler $U_t$.


A Clifford unitary $U_{0}$ on the total system of $n$ qubits sends Pauli strings to Pauli strings. As we dope the unitary by $t$ non-Clifford resources, $U_t$ will not generally send every Pauli string in another Pauli string. However, $U_t$ may still behave like a Clifford operation just on a subset of the Pauli group. To be precise, let $\mathbb{P}_{\Lambda}$ denote the Pauli group on the subsystem $\Lambda$, and let $G_{\Lambda}(U_t)$ denote such a preserved subset of the Pauli group, i.e., $G_{\Lambda}(U_t):=\{P\in\mathbb{P}_{\Lambda}\,|\, U_{t}^{\dag}PU_{t}\in\mathbb{P}_n\}$. In~\cite{leone2022LearningEfficientDecoders}, we prove that $|G_{\Lambda}(U_t)|\ge 2^{2|\Lambda|-t}$, i.e., a fraction of $2^{-t}$ Pauli operators are preserved by the action of $U_t$.

As we show in~\cite{leone2022LearningEfficientDecoders}, for any \textit{$t$-doped Clifford circuit} there exists a Clifford operation $\mathcal{D}$ and two subsystems $E_{1},E_{2}\subset D$ such that $\mathbb{P}_{E_1}\subseteq \mathcal{D}^{\dag}G_{D}(U_t)\mathcal{D}\subseteq \mathbb{P}_{E_2}$. However, for the purpose of this Letter, we work in the following simplified setting: assume that there exists a subset $E$ of qubits $E\subset D$ and a Clifford operation $\mathcal{D}\,:\,D\mapsto E\cup F$ such that $\mathcal{D}^{\dag}G_{D}(U_t)\mathcal{D}=\mathbb{P}_E$. In words, the diagonalizer $\mathcal{D}$ moves the non-Cliffordness around the subsystem $D$ and concentrates it all into the subsystem $F\equiv D\setminus E$. The simplified scenario described above corresponds to a special class of circuits in which $E_1\equiv E_2$, see~\footnote{The simplified settings of this Letter correspond to a special class of $t$-doped Clifford circuits. In particular, one can always write $U_t=C_0 W_{1}W_{2}\cdots W_t C_{0}^{\prime}$, where $C_{0},C_{0}^{\prime}$ are Clifford, while $W_{j}$ are $1$-doped Clifford circuits. The setting explored in this Letter corresponds to the special case where $[W_{j},W_{k}]=0$ for $j,k\in{1,t/2}$, and $W_{j+t/2}=C_{j}^{\dagger}W_{j}C_{j}$ for $j\in{1,\ldots,t/2}$, such that $[W_{j+t/2},W_{j}]\neq 0$, with ${C_{j}}$ being a set of commuting Clifford operators. In this manner, if $P$ is the Pauli generator not preserved by the action of $W_{j}$, then $C_{j}^{\dagger}PC_{j}$ is also not preserved.}.

By measuring the output subsystem $E$ only, the unitary operation $\mathcal{D}^{\dag}U_t$ is indistinguishable from a Clifford operator: the action on any Pauli string $P_E\in\mathbb{P}_E$ is, by construction, a Pauli string in $\mathbb{P}_n$
\be
(\mathcal{D}^{\dag}U_t)^{\dag}P_{E}(\mathcal{D}^{\dag}U_t)\in\mathbb{P}_n\label{indistinguishability}
\ee
By applying the diagonalizer, Bob effectively splits the output subsystem $D$ into two parts: $E$ that contains only Clifford information, and its complement $F$ that contains all the non-Cliffordness of $U_t$. In what follows, let us denote the adjoint action of $\mathcal{D}^{\dag}U_t$ on Pauli operators $P$ as $\tilde{P}\equiv(\mathcal{D}^{\dag}U_t)^{\dag}P (\mathcal{D}^{\dag}U_t)$ to lighten the notation. 

The question is this: can Bob only look at the subsystem $E$ and learn the information scrambled by $U_t$ by employing a Clifford decoder? The decoupling theorem says that the joint subsystem $E\cup B^{\prime}$ contains all the information (up to an error $\epsilon$) about $R$, provided that $|E|= |A|+\log\epsilon^{-1/2}$. Therefore, the answer is Yes, provided that Bob looks only at the joint subsystem $E\cup B^{\prime}$ by tracing out the subsystem $F$. In practice, this operation is equivalent to projecting onto the $\ket{EE^{\prime}}$ instead of $\ket{DD^{\prime}}$, and can be achieved by applying a randomizer, i.e., a Clifford operation $\mathcal{R}$ that, acting on $F^{\prime}\cup C^{\prime}$, scrambles the unwanted information contained in $F^{\prime}$ throughout the system $C^{\prime}$. The action of the randomizer results in hiding the non-Clifford information contained in $F$ in the subsystem $C$, which is equivalent to tracing out the subsystem $F$ with probability governed by the size of $C$. In practice, a randomizer $\mathcal{R}$ is just a random Clifford operator, which is scrambling with overwhelming probability~\cite{roberts2017ChaosComplexityDesign}.

At this point, a decrypter $V^{\prime}$ that reads the \textit{clean} Clifford information out of the subsystem $E$ is sufficient to completely decode the information in a Clifford-like fashion. $V^{\prime}$ is a Clifford unitary operator that obeys the following property: 
\be
(\mathcal{D}^{\dag}V^{\prime})^{\dag}P_E(\mathcal{D}^{\dag}V^{\prime})=\tilde{P}_E\quad\forall P_E\in\mathbb{P}_E
\label{decrypter}
\ee
i.e., it mimics the (Clifford-like) action~\eqref{indistinguishability} of the operator $\mathcal{D}^{\dag}U_t$ only on the local Pauli group $\mathbb{P}_E$. The reason behind this capability is that the unitary operator $\mathcal{D}^{\dag}U_t$ is practically indistinguishable from a Clifford operator from the point of view of an observer that measures the subsystem $E$ only, see Eq.~\eqref{indistinguishability}. Let us denote the adjoint action of $\mathcal{D}^{\dag}V^{\prime}$ on $P\in\mathbb{P}$ as $\hat{P}\equiv(\mathcal{D}^{\dag}V^{\prime})^{\dag}P(\mathcal{D}^{\dag}V^{\prime})$. 

In summary, the decoding protocol consists of the following steps starting from Eq.~\eqref{statescrambled}: (i) apply the diagonalizer $\mathcal{D}$ on the output $D$ of the scrambler $U_t$; (ii) append $\ket{A^{\prime}R^{\prime}}$; (iii) apply the decrypter $V^{\prime T}$ followed by $\mathcal{D}^{*}$ (the star denotes the conjugate operation); (iv) apply the randomizer $\mathcal{R}^{*}$ on $F^{\prime}\cup C^{\prime}$, and (v) project onto $\ket{DD^{\prime}}$. The final state $\ket{\Psi_V}$ with $V=\mathcal{D}\mathcal{R}\mathcal{D}^{\dag}V^{\prime}$ can be represented diagrammatically as in Fig.~\ref{fig:diagram}.

We are just left to show that the decoder $V$ built as described above achieves the promised fidelity in Eq.~\eqref{mainF}. It is possible to show~\cite{SeeSupplementalMaterial} that for $V=\mathcal{D}\mathcal{R}\mathcal{D}^{\dag}V^{\prime}$, the fidelity reads
\be
\mathcal{F}(V)=\frac{1}{4^{|A|}}\frac{\braket{\tr(\tilde{P}_{D}\mathcal{R}^{\dag}\hat{P}_{D}\mathcal{R})}_{D}}{\braket{\tr(P_A\tilde{P}_{D}P_A\mathcal{R}^{\dag}\hat{P}_{D}\mathcal{R})}_{D,A}}
\label{fidelityprerandomizer}
\ee
We recall that $\tilde{P},\hat{P}$ represents the adjoint action of $\mathcal{D}^{\dag}U_t$ and $\mathcal{D}^{\dag}V^{\prime}$ on $P$, respectively. From Eq.~\eqref{fidelityprerandomizer}, it is possible to see that selecting the randomizer $\mathcal{R}$ as a random Clifford operator is equivalent to tracing out the unwanted non-Clifford information. Indeed, in the Supplemental Material~\cite{SeeSupplementalMaterial}, we show that, with failure probability $O(2^{-2|C|})$~\footnote{Note that this probability is different from the one displayed in Eq.~\eqref{successprobability}. The reason is that Čebyšëv inequality allows us to say that $\mathcal{F}(V)=\frac{1}{1+2^{2|A|-2|E|}}+O(\varepsilon)$ with probability $O(\varepsilon^{-2}2^{-2|C|})$, while the more detailed analysis of~\cite{leone2022LearningEfficientDecoders} shows $\mathcal{F}(V)\ge\frac{1}{1+2^{2|A|-2|E|}}$ (see Eq.~\eqref{mainF}) with probability given by Eq.~\eqref{successprobability}.} in the choice of the randomizer $\mathcal{R}$, the fidelity reads
\begin{equation}
\mathcal{F}(V)\simeq\frac{1}{4^{|A|}}\frac{\braket{\tr(\tilde{P}_{E}\hat{P}_{E})}_{E}}{\braket{\tr(P_A\tilde{P}_{E}P_A\hat{P}_{E})}_{E,A}}\,.
\label{fidelityE}
\end{equation}
The above equation tells us that the action of the randomizer is equivalent to tracing out the $F\equiv D\setminus E$ subsystem:  the average $\braket{\cdot}_{E}$ is now restricted on $E$ only. 

From the definition of the decrypter in Eq.~\eqref{decrypter}, one has that $\hat{P}_E=\tilde{P}_{E}$, therefore $\tr(\tilde{P}_E\hat{P}_E)=2^{n}$ and by definition in Eq.~\eqref{otocdef} one has $2^{-n}\braket{\tr(P_A\tilde{P}_{E}P_A\hat{P}_{E})}_{A,E}=\Omega_{AE}(U_t)$ . Therefore, from Eq.~\eqref{fidelityE}, and using Eq.~\eqref{scramblingdef} for  $U_t$ being scrambling, we obtain the following value for the fidelity
\begin{equation}
\mathcal{F}(V)\simeq\frac{1}{1+2^{2|A|-2|E|}}\label{fidelityrec}\,.
\end{equation}
The above equation shows that, by employing the diagonalizer and a randomizer, Bob is able to distill an EPR pair between $R$ and $R^{\prime}$ with fidelity approaching one exponentially fast in $|E|-|A|$. However, there is an important caveat: the size of the subsystem $E$ does depend on the number $t$ of non-Clifford gates. For $t=0$, $E=D$. More generally, we are only assured that $|E|\ge |D|-t/2$, recovering Eq.~\eqref{mainF}. If we allow an $\epsilon$ error for the fidelity, we have the following condition
$
|D|\ge |A|+t/2+\log\epsilon^{-1}
$, 
i.e., to make a constant error $\epsilon$, Bob must collect a linearly increasing (in $t$) number of output qubits $D$, rendering the unscrambling process increasingly nonlocal. 
At the same time, Eq.~\eqref{mainP} shows that the success probability shrinks with $|D|$. This is because the randomizer becomes less effective in scrambling if $C$ is small and $|C|=n-|D|$. The algorithm breaks down as $t\sim n$, as both the probability of learning and the fidelity of recovery start decaying exponentially in $n$.

\prlsection{Learning the Clifford decoder}
 The Clifford decoder~\eqref{thedecoder} is capable of decoding the input information scrambled by the $t$-doped Clifford circuit $U_t$. In this section, we show how one can learn each component $\mathcal{D},\mathcal{R},V^{\prime}$ of the Clifford decoder $V$ by observing the output subsystem $D$. Specifically, we assume black-box access to $U_t$, meaning we can apply one or multiple copies of $U_t$ on a quantum register and measure the output $D$. We note that the ability to construct the decoder $V$ solely by analyzing the output $D$ is desirable in contexts in which the output subsystem $C$ is inaccessible to the observer~\cite{hayden2007BlackHolesMirrors}.


While the rigorous version of the learning algorithm needs to be found in~\cite{leone2022LearningEfficientDecoders}, here we highlight the key steps of the learning process. As we said, given the output subsystem $D$, there exists a subgroup $G_{D}(U_t)$ consisting of Pauli strings that are sent to Pauli strings by the adjoint action of $U_t$. By employing entangling measurements (yet stabilizer) on the output of $U_t$ on test Pauli operators $P_D\in\mathbb{P}_D$, one can decide whether $U_t^{\dag}P_DU_t$ is a Pauli string or not. By repeating this procedure multiple times for different test Pauli operators, one can effectively learn a set of generators $g_D(U_t)$ for the group $G_{D}(U_t)$. Notice that the above procedure reveals also the image group, denoted as $U_{t}^{\dag}G_{D}(U_t)U_t$, of $G_{D}(U_t)$ through the adjoint action of $U_t$. The total time and query complexity of the algorithm is polynomial in $n$ while exponential in the number of non-Clifford gates $t$.  

At this point, having learned the groups $G_{D}(U_t)$ and $U_{t}^{\dag}G_{D}(U_t)U_t$, one can construct the Clifford decoder $V$ through classical postprocessing in polynomial time. To build the diagonalizer $\mathcal{D}$, it is sufficient, through manipulation of the tableau representation of Clifford circuits~\cite{aaronson2004ImprovedSimulationStabilizer}, to construct a Clifford operation that sends $G_{D}(U_t)$ to a Pauli group $\mathbb{P}_E$ on a subsystem $E$ of size $|E|=\log|G_{D}(U_t)|$.  As the swap operator between qubits belongs to the Clifford group, the learner can freely choose the subsystem $E$ among the qubits in $D$. Similarly, it is possible to distill the decrypter $V^{\prime}$ that sends $G_{D}(U_t)$ to $U_{t}^{\dag}G_{D}(U_t)U_t\equiv (\mathcal{D}^{\dag}U_t)^{\dag}\mathbb{P}_E(\mathcal{D}^{\dag}U_t)$. Notice that existence of both $\mathcal{D}$ and $V^{\prime}$ is guaranteed by the Gottesman-Knill theorem~\cite{gottesman1998HeisenbergRepresentationQuantuma}. Finally, to build the randomizer $\mathcal{R}$, it is sufficient to draw a Clifford operator uniformly at random. As a result, Bob is able to construct the Clifford decoder $V$ by having black-box access to the scrambler $U_t$ and reading the output $D$ of $U_t$ in time $\poly(n,2^t)$. 

\prlsection{Black-hole scrambling and Clifford decoding}
In this section, we provide a brief overview of the potential implications of our findings in the context of black hole physics. Black holes are inherently isolated objects and thus, if the laws of quantum mechanics hold, their internal dynamics must be unitary. Traditionally, black holes have been conceptualized as exhibiting maximally chaotic unitary behavior~\cite{harlow2023blackholesquantumgravity,yang2023ComplexityLearningPseudo}, being often described as random unitary operators~\cite{page1993AverageEntropySubsystem}, to account for their intrinsic complexity. Furthermore, black holes are widely believed to be the fastest scramblers in nature~\cite{sekino2008FastScramblers}.

However, the assumption of resembling random unitary dynamics for fast scrambling might be overly stringent. This is because, as discussed above, Clifford circuits excel at efficiently scrambling information. We can thus challenge the conventional notion of characterizing black holes as maximally chaotic systems and instead focus solely on their (fast) scrambling properties. Under this perspective, the internal dynamics of a black hole could potentially be described by Clifford circuits or, more generally, by $t$-doped Clifford circuits, which are increasingly more chaotic with $t$\cite{leone2021QuantumChaosQuantum}.

If this hypothesis holds, our results carry significant implications. Specifically, they suggest that information $A$ entering a black hole and subsequently expelled through Hawking radiation $D$ could be effectively learned by an observer employing a Clifford decoder. By introducing test information into the black hole and analyzing the outgoing Hawking radiation, Bob could learn how to decode the information contained in the radiation emitted by a black hole, without accessing the black hole interior $C$. This decoder could then be employed to investigate the physics in the proximity of the black hole.

\prlsection{Conclusions}
Complex quantum operations require an exponential number of classical resources to be represented and simulated. However, important properties of complex (but not fully chaotic) quantum operations---like the decoding of scrambled information from Hawking radiation---can be both learned and simulated efficiently in a classical computer by pushing the complex behavior residing in the non-Clifford resources to noisy subsystems. We speculate this behavior can be extended to the general framework of quantum error-correcting codes. A practical direction to pursue for future research is to investigate the robustness of Clifford decoding in the presence of noise.  If we see the scrambling of the information tossed in the scrambler as a quantum process, all its efficient decoders are equivalent in characterizing it. It would thus be interesting to see how the algorithm proposed here can be expanded to improve process tomography.

\prlsection{Acknowledgments}
The authors thank C. Chamon, B. Yan, D. Lewis, and A. Scocco for important discussions. S.F.E.O. and L.L. acknowledge support from NSF grant no. 2014000. A.H. acknowledges financial support from PNRR MUR project PE0000023-NQSTI and PNRR MUR project CN 00000013-ICSC. The work of L.L. and S.F.E.O. was supported in part by the U.S. Department of Energy (DOE) through a quantum computing program sponsored by the Los Alamos National Laboratory Information Science and Technology Institute, and by the  Center for Nonlinear Studies at Los Alamos National Laboratory (LANL). S.F.E.O. and L.L. contributed equally to this work.

\let\oldaddcontentsline\addcontentsline
\renewcommand{\addcontentsline}[3]{}
\medskip


%


\begin{thebibliography}{43}%
\makeatletter
\providecommand \@ifxundefined [1]{%
 \@ifx{#1\undefined}
}%
\providecommand \@ifnum [1]{%
 \ifnum #1\expandafter \@firstoftwo
 \else \expandafter \@secondoftwo
 \fi
}%
\providecommand \@ifx [1]{%
 \ifx #1\expandafter \@firstoftwo
 \else \expandafter \@secondoftwo
 \fi
}%
\providecommand \natexlab [1]{#1}%
\providecommand \enquote  [1]{``#1''}%
\providecommand \bibnamefont  [1]{#1}%
\providecommand \bibfnamefont [1]{#1}%
\providecommand \citenamefont [1]{#1}%
\providecommand \href@noop [0]{\@secondoftwo}%
\providecommand \href [0]{\begingroup \@sanitize@url \@href}%
\providecommand \@href[1]{\@@startlink{#1}\@@href}%
\providecommand \@@href[1]{\endgroup#1\@@endlink}%
\providecommand \@sanitize@url [0]{\catcode `\\12\catcode `\$12\catcode
  `\&12\catcode `\#12\catcode `\^12\catcode `\_12\catcode `\%12\relax}%
\providecommand \@@startlink[1]{}%
\providecommand \@@endlink[0]{}%
\providecommand \url  [0]{\begingroup\@sanitize@url \@url }%
\providecommand \@url [1]{\endgroup\@href {#1}{\urlprefix }}%
\providecommand \urlprefix  [0]{URL }%
\providecommand \Eprint [0]{\href }%
\providecommand \doibase [0]{http://dx.doi.org/}%
\providecommand \selectlanguage [0]{\@gobble}%
\providecommand \bibinfo  [0]{\@secondoftwo}%
\providecommand \bibfield  [0]{\@secondoftwo}%
\providecommand \translation [1]{[#1]}%
\providecommand \BibitemOpen [0]{}%
\providecommand \bibitemStop [0]{}%
\providecommand \bibitemNoStop [0]{.\EOS\space}%
\providecommand \EOS [0]{\spacefactor3000\relax}%
\providecommand \BibitemShut  [1]{\csname bibitem#1\endcsname}%
\let\auto@bib@innerbib\@empty
\bibitem [{Note1()}]{Note1}%
  \BibitemOpen
  \bibinfo {note} {Humpty Dumpty sat on a wall, Humpty Dumpty had a great
  fall. All the king's horses and all the king's men. Couldn't put Humpty together again.}\BibitemShut {Stop}%
\bibitem [{\citenamefont {Hosur}\ \emph {et~al.}(2016)\citenamefont {Hosur},
  \citenamefont {Qi}, \citenamefont {Roberts},\ and\ \citenamefont
  {Yoshida}}]{hosur2016ChaosQuantumChannels}%
  \BibitemOpen
  \bibfield  {author} {\bibinfo {author} {\bibfnamefont {P.}~\bibnamefont
  {Hosur}}, \bibinfo {author} {\bibfnamefont {X.-L.}\ \bibnamefont {Qi}},
  \bibinfo {author} {\bibfnamefont {D.~A.}\ \bibnamefont {Roberts}}, \ and\
  \bibinfo {author} {\bibfnamefont {B.}~\bibnamefont {Yoshida}},\ }\href
  {\doibase 10.1007/JHEP02(2016)004} {\bibfield  {journal} {\bibinfo  {journal}
  {J. High Energy Phys.}\ }\textbf {\bibinfo {volume} {2016}},\ \bibinfo
  {pages} {4} (\bibinfo {year} {2016})}\BibitemShut {NoStop}%
\bibitem [{\citenamefont {Ding}\ \emph {et~al.}(2016)\citenamefont {Ding},
  \citenamefont {Hayden},\ and\ \citenamefont
  {Walter}}]{ding2016ConditionalMutualInformation}%
  \BibitemOpen
  \bibfield  {author} {\bibinfo {author} {\bibfnamefont {D.}~\bibnamefont
  {Ding}}, \bibinfo {author} {\bibfnamefont {P.}~\bibnamefont {Hayden}}, \ and\
  \bibinfo {author} {\bibfnamefont {M.}~\bibnamefont {Walter}},\ }\href
  {\doibase 10.1007/JHEP12(2016)145} {\bibfield  {journal} {\bibinfo  {journal}
  {J. High Energy Phys.}\ }\textbf {\bibinfo {volume} {2016}},\ \bibinfo
  {pages} {145} (\bibinfo {year} {2016})}\BibitemShut {NoStop}%
\bibitem [{\citenamefont {Brown}\ and\ \citenamefont
  {Fawzi}(2013)}]{brown2013ScramblingSpeedRandom}%
  \BibitemOpen
  \bibfield  {author} {\bibinfo {author} {\bibfnamefont {W.}~\bibnamefont
  {Brown}}\ and\ \bibinfo {author} {\bibfnamefont {O.}~\bibnamefont {Fawzi}},\
  }\href {\doibase 10.48550/arXiv.1210.6644} {\enquote {\bibinfo {title}
  {Scrambling speed of random quantum circuits},}\ } (\bibinfo {year} {2013}),\
  \Eprint {http://arxiv.org/abs/1210.6644} {arXiv:1210.6644 [hep-th,
  physics:quant-ph]} \BibitemShut {NoStop}%
\bibitem [{\citenamefont {Liu}\ \emph {et~al.}(2018{\natexlab{a}})\citenamefont
  {Liu}, \citenamefont {Lloyd}, \citenamefont {Zhu},\ and\ \citenamefont
  {Zhu}}]{liu2018GeneralizedEntanglementEntropies}%
  \BibitemOpen
  \bibfield  {author} {\bibinfo {author} {\bibfnamefont {Z.-W.}\ \bibnamefont
  {Liu}}, \bibinfo {author} {\bibfnamefont {S.}~\bibnamefont {Lloyd}}, \bibinfo
  {author} {\bibfnamefont {E.~Y.}\ \bibnamefont {Zhu}}, \ and\ \bibinfo
  {author} {\bibfnamefont {H.}~\bibnamefont {Zhu}},\ }\href {\doibase
  10.1103/PhysRevLett.120.130502} {\bibfield  {journal} {\bibinfo  {journal}
  {Phys. Rev. Lett,}\ }\textbf {\bibinfo {volume} {120}},\ \bibinfo {pages}
  {130502} (\bibinfo {year} {2018}{\natexlab{a}})}\BibitemShut {NoStop}%
\bibitem [{\citenamefont {Liu}\ \emph {et~al.}(2018{\natexlab{b}})\citenamefont
  {Liu}, \citenamefont {Lloyd}, \citenamefont {Zhu},\ and\ \citenamefont
  {Zhu}}]{liu2018EntanglementQuantumRandomness}%
  \BibitemOpen
  \bibfield  {author} {\bibinfo {author} {\bibfnamefont {Z.-W.}\ \bibnamefont
  {Liu}}, \bibinfo {author} {\bibfnamefont {S.}~\bibnamefont {Lloyd}}, \bibinfo
  {author} {\bibfnamefont {E.}~\bibnamefont {Zhu}}, \ and\ \bibinfo {author}
  {\bibfnamefont {H.}~\bibnamefont {Zhu}},\ }\href {\doibase
  10.1007/JHEP07(2018)041} {\bibfield  {journal} {\bibinfo  {journal} {J. High
  Energy Phys.}\ }\textbf {\bibinfo {volume} {2018}},\ \bibinfo {pages} {41}
  (\bibinfo {year} {2018}{\natexlab{b}})}\BibitemShut {NoStop}%
\bibitem [{\citenamefont {Styliaris}\ \emph {et~al.}(2021)\citenamefont
  {Styliaris}, \citenamefont {Anand},\ and\ \citenamefont
  {Zanardi}}]{styliaris2021InformationScramblingBipartitions}%
  \BibitemOpen
  \bibfield  {author} {\bibinfo {author} {\bibfnamefont {G.}~\bibnamefont
  {Styliaris}}, \bibinfo {author} {\bibfnamefont {N.}~\bibnamefont {Anand}}, \
  and\ \bibinfo {author} {\bibfnamefont {P.}~\bibnamefont {Zanardi}},\ }\href
  {\doibase 10.1103/PhysRevLett.126.030601} {\bibfield  {journal} {\bibinfo
  {journal} {Phys. Rev. Lett,}\ }\textbf {\bibinfo {volume} {126}},\ \bibinfo
  {pages} {030601} (\bibinfo {year} {2021})}\BibitemShut {NoStop}%
\bibitem [{\citenamefont {Lloyd}(1988)}]{lloyd1988BlackHolesDemons}%
  \BibitemOpen
  \bibfield  {author} {\bibinfo {author} {\bibfnamefont {S.}~\bibnamefont
  {Lloyd}},\ }\emph {\bibinfo {title} {Black {{Holes}}, {{Demons}} and the
  {{Loss}} of {{Coherence}}: {{How}} Complex Systems Get Information, and What
  They Do with It}},\ \href@noop {} {Ph.D. thesis},\ \bibinfo  {school}
  {Rockefeller University} (\bibinfo {year} {1988})\BibitemShut {NoStop}%
\bibitem [{\citenamefont
  {Page}(1993{\natexlab{a}})}]{page1993InformationBlackHole}%
  \BibitemOpen
  \bibfield  {author} {\bibinfo {author} {\bibfnamefont {D.~N.}\ \bibnamefont
  {Page}},\ }\href {\doibase 10.1103/PhysRevLett.71.3743} {\bibfield  {journal}
  {\bibinfo  {journal} {Phys. Rev. Lett,}\ }\textbf {\bibinfo {volume} {71}},\
  \bibinfo {pages} {3743} (\bibinfo {year} {1993}{\natexlab{a}})}\BibitemShut
  {NoStop}%
\bibitem [{\citenamefont {Hayden}\ and\ \citenamefont
  {Preskill}(2007)}]{hayden2007BlackHolesMirrors}%
  \BibitemOpen
  \bibfield  {author} {\bibinfo {author} {\bibfnamefont {P.}~\bibnamefont
  {Hayden}}\ and\ \bibinfo {author} {\bibfnamefont {J.}~\bibnamefont
  {Preskill}},\ }\href {\doibase 10.1088/1126-6708/2007/09/120} {\bibfield
  {journal} {\bibinfo  {journal} {J. High Energy Phys.}\ }\textbf {\bibinfo
  {volume} {2007}},\ \bibinfo {pages} {120} (\bibinfo {year}
  {2007})}\BibitemShut {NoStop}%
\bibitem [{\citenamefont {Shenker}\ and\ \citenamefont
  {Stanford}(2014{\natexlab{a}})}]{shenker2014BlackHolesButterfly}%
  \BibitemOpen
  \bibfield  {author} {\bibinfo {author} {\bibfnamefont {S.~H.}\ \bibnamefont
  {Shenker}}\ and\ \bibinfo {author} {\bibfnamefont {D.}~\bibnamefont
  {Stanford}},\ }\href {\doibase 10.1007/JHEP03(2014)067} {\bibfield  {journal}
  {\bibinfo  {journal} {J. High Energy Phys.}\ }\textbf {\bibinfo {volume}
  {2014}},\ \bibinfo {pages} {67} (\bibinfo {year}
  {2014}{\natexlab{a}})}\BibitemShut {NoStop}%
\bibitem [{\citenamefont {Shenker}\ and\ \citenamefont
  {Stanford}(2015)}]{shenker2015StringyEffectsScrambling}%
  \BibitemOpen
  \bibfield  {author} {\bibinfo {author} {\bibfnamefont {S.~H.}\ \bibnamefont
  {Shenker}}\ and\ \bibinfo {author} {\bibfnamefont {D.}~\bibnamefont
  {Stanford}},\ }\href {\doibase 10.1007/JHEP05(2015)132} {\bibfield  {journal}
  {\bibinfo  {journal} {J. High Energy Phys.}\ }\textbf {\bibinfo {volume}
  {2015}},\ \bibinfo {pages} {132} (\bibinfo {year} {2015})}\BibitemShut
  {NoStop}%
\bibitem [{\citenamefont {Roberts}\ \emph {et~al.}(2015)\citenamefont
  {Roberts}, \citenamefont {Stanford},\ and\ \citenamefont
  {Susskind}}]{roberts2015LocalizedShocks}%
  \BibitemOpen
  \bibfield  {author} {\bibinfo {author} {\bibfnamefont {D.~A.}\ \bibnamefont
  {Roberts}}, \bibinfo {author} {\bibfnamefont {D.}~\bibnamefont {Stanford}}, \
  and\ \bibinfo {author} {\bibfnamefont {L.}~\bibnamefont {Susskind}},\ }\href
  {\doibase 10.1007/JHEP03(2015)051} {\bibfield  {journal} {\bibinfo  {journal}
  {J. High Energy Phys.}\ }\textbf {\bibinfo {volume} {2015}},\ \bibinfo
  {pages} {51} (\bibinfo {year} {2015})}\BibitemShut {NoStop}%
\bibitem [{\citenamefont {Shenker}\ and\ \citenamefont
  {Stanford}(2014{\natexlab{b}})}]{shenker2014MultipleShocks}%
  \BibitemOpen
  \bibfield  {author} {\bibinfo {author} {\bibfnamefont {S.~H.}\ \bibnamefont
  {Shenker}}\ and\ \bibinfo {author} {\bibfnamefont {D.}~\bibnamefont
  {Stanford}},\ }\href {\doibase 10.1007/JHEP12(2014)046} {\bibfield  {journal}
  {\bibinfo  {journal} {J. High Energy Phys.}\ }\textbf {\bibinfo {volume}
  {2014}},\ \bibinfo {pages} {46} (\bibinfo {year}
  {2014}{\natexlab{b}})}\BibitemShut {NoStop}%
\bibitem [{\citenamefont {Sekino}\ and\ \citenamefont
  {Susskind}(2008)}]{sekino2008FastScramblers}%
  \BibitemOpen
  \bibfield  {author} {\bibinfo {author} {\bibfnamefont {Y.}~\bibnamefont
  {Sekino}}\ and\ \bibinfo {author} {\bibfnamefont {L.}~\bibnamefont
  {Susskind}},\ }\href {\doibase 10.1088/1126-6708/2008/10/065} {\bibfield
  {journal} {\bibinfo  {journal} {J. High Energy Phys.}\ }\textbf {\bibinfo
  {volume} {2008}},\ \bibinfo {pages} {065} (\bibinfo {year}
  {2008})}\BibitemShut {NoStop}%
\bibitem [{\citenamefont {Lashkari}\ \emph {et~al.}(2013)\citenamefont
  {Lashkari}, \citenamefont {Stanford}, \citenamefont {Hastings}, \citenamefont
  {Osborne},\ and\ \citenamefont
  {Hayden}}]{lashkari2013FastScramblingConjecture}%
  \BibitemOpen
  \bibfield  {author} {\bibinfo {author} {\bibfnamefont {N.}~\bibnamefont
  {Lashkari}}, \bibinfo {author} {\bibfnamefont {D.}~\bibnamefont {Stanford}},
  \bibinfo {author} {\bibfnamefont {M.}~\bibnamefont {Hastings}}, \bibinfo
  {author} {\bibfnamefont {T.}~\bibnamefont {Osborne}}, \ and\ \bibinfo
  {author} {\bibfnamefont {P.}~\bibnamefont {Hayden}},\ }\href {\doibase
  10.1007/JHEP04(2013)022} {\bibfield  {journal} {\bibinfo  {journal} {J. High
  Energy Phys.}\ }\textbf {\bibinfo {volume} {2013}},\ \bibinfo {pages} {22}
  (\bibinfo {year} {2013})}\BibitemShut {NoStop}%
\bibitem [{\citenamefont {Maldacena}\ \emph {et~al.}(2016)\citenamefont
  {Maldacena}, \citenamefont {Shenker},\ and\ \citenamefont
  {Stanford}}]{maldacena2016BoundChaos}%
  \BibitemOpen
  \bibfield  {author} {\bibinfo {author} {\bibfnamefont {J.}~\bibnamefont
  {Maldacena}}, \bibinfo {author} {\bibfnamefont {S.~H.}\ \bibnamefont
  {Shenker}}, \ and\ \bibinfo {author} {\bibfnamefont {D.}~\bibnamefont
  {Stanford}},\ }\href {\doibase 10.1007/JHEP08(2016)106} {\bibfield  {journal}
  {\bibinfo  {journal} {J. High Energy Phys.}\ }\textbf {\bibinfo {volume}
  {2016}},\ \bibinfo {pages} {106} (\bibinfo {year} {2016})}\BibitemShut
  {NoStop}%
\bibitem [{\citenamefont {Kitaev}(2014)}]{kitaev2014HiddenCorrelationsHawking}%
  \BibitemOpen
  \bibfield  {author} {\bibinfo {author} {\bibfnamefont {A.}~\bibnamefont
  {Kitaev}},\ }in\ \href {https://online.kitp.ucsb.edu/online/joint98/kitaev/}
  {\emph {\bibinfo {booktitle} {{P}roceedings of the Fundamental Physics Prize Symposium}}} (\bibinfo {year}
  {2014}),\ Vol.~\bibinfo {volume} {10}\ \BibitemShut {NoStop}%
\bibitem [{\citenamefont {Roberts}\ and\ \citenamefont
  {Swingle}(2016)}]{roberts2016LiebRobinsonBoundButterfly}%
  \BibitemOpen
  \bibfield  {author} {\bibinfo {author} {\bibfnamefont {D.~A.}\ \bibnamefont
  {Roberts}}\ and\ \bibinfo {author} {\bibfnamefont {B.}~\bibnamefont
  {Swingle}},\ }\href {\doibase 10.1103/PhysRevLett.117.091602} {\bibfield
  {journal} {\bibinfo  {journal} {Phys. Rev. Lett,}\ }\textbf {\bibinfo
  {volume} {117}},\ \bibinfo {pages} {091602} (\bibinfo {year}
  {2016})}\BibitemShut {NoStop}%
\bibitem [{\citenamefont {Zhou}\ \emph {et~al.}(2020)\citenamefont {Zhou},
  \citenamefont {Yang}, \citenamefont {Hamma},\ and\ \citenamefont
  {Chamon}}]{zhou2020SingleGateClifford}%
  \BibitemOpen
  \bibfield  {author} {\bibinfo {author} {\bibfnamefont {S.}~\bibnamefont
  {Zhou}}, \bibinfo {author} {\bibfnamefont {Z.-C.}\ \bibnamefont {Yang}},
  \bibinfo {author} {\bibfnamefont {A.}~\bibnamefont {Hamma}}, \ and\ \bibinfo
  {author} {\bibfnamefont {C.}~\bibnamefont {Chamon}},\ }\href {\doibase
  10.21468/SciPostPhys.9.6.087} {\bibfield  {journal} {\bibinfo  {journal}
  {SciPost Phys.}\ }\textbf {\bibinfo {volume} {9}},\ \bibinfo {pages} {87}
  (\bibinfo {year} {2020})}\BibitemShut {NoStop}%
\bibitem [{\citenamefont {Leone}\ \emph
  {et~al.}(2021{\natexlab{a}})\citenamefont {Leone}, \citenamefont {Oliviero},
  \citenamefont {Zhou},\ and\ \citenamefont
  {Hamma}}]{leone2021QuantumChaosQuantum}%
  \BibitemOpen
  \bibfield  {author} {\bibinfo {author} {\bibfnamefont {L.}~\bibnamefont
  {Leone}}, \bibinfo {author} {\bibfnamefont {S.~F.~E.}\ \bibnamefont
  {Oliviero}}, \bibinfo {author} {\bibfnamefont {Y.}~\bibnamefont {Zhou}}, \
  and\ \bibinfo {author} {\bibfnamefont {A.}~\bibnamefont {Hamma}},\ }\href
  {\doibase 10.22331/q-2021-05-04-453} {\bibfield  {journal} {\bibinfo
  {journal} {Quantum}\ }\textbf {\bibinfo {volume} {5}},\ \bibinfo {pages}
  {453} (\bibinfo {year} {2021}{\natexlab{a}})}\BibitemShut {NoStop}%
\bibitem [{\citenamefont {Oliviero}\ \emph
  {et~al.}(2021{\natexlab{a}})\citenamefont {Oliviero}, \citenamefont {Leone},\
  and\ \citenamefont {Hamma}}]{oliviero2021TransitionsEntanglementComplexity}%
  \BibitemOpen
  \bibfield  {author} {\bibinfo {author} {\bibfnamefont {S.~F.~E.}\
  \bibnamefont {Oliviero}}, \bibinfo {author} {\bibfnamefont {L.}~\bibnamefont
  {Leone}}, \ and\ \bibinfo {author} {\bibfnamefont {A.}~\bibnamefont
  {Hamma}},\ }\href {\doibase 10.1016/j.physleta.2021.127721} {\bibfield
  {journal} {\bibinfo  {journal} {Phys. Lett. A}\ }\textbf {\bibinfo {volume}
  {418}},\ \bibinfo {pages} {127721} (\bibinfo {year}
  {2021}{\natexlab{a}})}\BibitemShut {NoStop}%
\bibitem [{\citenamefont {True}\ and\ \citenamefont
  {Hamma}(2022)}]{true2022TransitionsEntanglementComplexity}%
  \BibitemOpen
  \bibfield  {author} {\bibinfo {author} {\bibfnamefont {S.}~\bibnamefont
  {True}}\ and\ \bibinfo {author} {\bibfnamefont {A.}~\bibnamefont {Hamma}},\
  }\href {\doibase 10.22331/q-2022-09-22-818} {\bibfield  {journal} {\bibinfo
  {journal} {Quantum}\ }\textbf {\bibinfo {volume} {6}},\ \bibinfo {pages}
  {818} (\bibinfo {year} {2022})}\BibitemShut {NoStop}%
\bibitem [{\citenamefont {Low}(2009)}]{low2009LearningTestingAlgorithms}%
  \BibitemOpen
  \bibfield  {author} {\bibinfo {author} {\bibfnamefont {R.~A.}\ \bibnamefont
  {Low}},\ }\href {\doibase 10.1103/PhysRevA.80.052314} {\bibfield  {journal}
  {\bibinfo  {journal} {Phys. Rev. A}\ }\textbf {\bibinfo {volume} {80}},\
  \bibinfo {pages} {052314} (\bibinfo {year} {2009})}\BibitemShut {NoStop}%
\bibitem [{\citenamefont {Lai}\ and\ \citenamefont
  {Cheng}(2022)}]{lai2022LearningQuantumCircuits}%
  \BibitemOpen
  \bibfield  {author} {\bibinfo {author} {\bibfnamefont {C.-Y.}\ \bibnamefont
  {Lai}}\ and\ \bibinfo {author} {\bibfnamefont {H.-C.}\ \bibnamefont
  {Cheng}},\ }\href {\doibase 10.1109/tit.2022.3151760} {\bibfield  {journal}
  {\bibinfo  {journal} {IEEE Trans. Inf. Theory}\ }\textbf {\bibinfo {volume}
  {68}},\ \bibinfo {pages} {3951} (\bibinfo {year} {2022})}\BibitemShut
  {NoStop}%
\bibitem [{\citenamefont {Chamon}\ \emph {et~al.}(2022)\citenamefont {Chamon},
  \citenamefont {Mucciolo},\ and\ \citenamefont
  {Ruckenstein}}]{chamon2022QuantumStatisticalMechanics}%
  \BibitemOpen
  \bibfield  {author} {\bibinfo {author} {\bibfnamefont {C.}~\bibnamefont
  {Chamon}}, \bibinfo {author} {\bibfnamefont {E.~R.}\ \bibnamefont
  {Mucciolo}}, \ and\ \bibinfo {author} {\bibfnamefont {A.~E.}\ \bibnamefont
  {Ruckenstein}},\ }\href {\doibase 10.1016/j.aop.2022.169086} {\bibfield
  {journal} {\bibinfo  {journal} {Ann. Phys.(Amsterdam)}\ }\textbf {\bibinfo
  {volume} {446}},\ \bibinfo {pages} {169086} (\bibinfo {year}
  {2022})}\BibitemShut {NoStop}%
\bibitem [{\citenamefont {Roberts}\ and\ \citenamefont
  {Yoshida}(2017)}]{roberts2017ChaosComplexityDesign}%
  \BibitemOpen
  \bibfield  {author} {\bibinfo {author} {\bibfnamefont {D.~A.}\ \bibnamefont
  {Roberts}}\ and\ \bibinfo {author} {\bibfnamefont {B.}~\bibnamefont
  {Yoshida}},\ }\href {\doibase 10.1007/JHEP04(2017)121} {\bibfield  {journal}
  {\bibinfo  {journal} {J. High Energy Phys.}\ }\textbf {\bibinfo {volume}
  {2017}},\ \bibinfo {pages} {121} (\bibinfo {year} {2017})}\BibitemShut
  {NoStop}%
\bibitem [{\citenamefont {Oliviero}\ \emph
  {et~al.}(2021{\natexlab{b}})\citenamefont {Oliviero}, \citenamefont {Leone},
  \citenamefont {Caravelli},\ and\ \citenamefont
  {Hamma}}]{oliviero2021RandomMatrixTheory}%
  \BibitemOpen
  \bibfield  {author} {\bibinfo {author} {\bibfnamefont {S.~F.~E.}\
  \bibnamefont {Oliviero}}, \bibinfo {author} {\bibfnamefont {L.}~\bibnamefont
  {Leone}}, \bibinfo {author} {\bibfnamefont {F.}~\bibnamefont {Caravelli}}, \
  and\ \bibinfo {author} {\bibfnamefont {A.}~\bibnamefont {Hamma}},\ }\href
  {\doibase 10.21468/SciPostPhys.10.3.076} {\bibfield  {journal} {\bibinfo
  {journal} {SciPost Phys.}\ }\textbf {\bibinfo {volume} {10}},\ \bibinfo
  {pages} {76} (\bibinfo {year} {2021}{\natexlab{b}})}\BibitemShut {NoStop}%
\bibitem [{\citenamefont {Leone}\ \emph
  {et~al.}(2021{\natexlab{b}})\citenamefont {Leone}, \citenamefont {Oliviero},\
  and\ \citenamefont {Hamma}}]{leone2021IsospectralTwirlingQuantum}%
  \BibitemOpen
  \bibfield  {author} {\bibinfo {author} {\bibfnamefont {L.}~\bibnamefont
  {Leone}}, \bibinfo {author} {\bibfnamefont {S.~F.~E.}\ \bibnamefont
  {Oliviero}}, \ and\ \bibinfo {author} {\bibfnamefont {A.}~\bibnamefont
  {Hamma}},\ }\href {\doibase 10.3390/e23081073} {\bibfield  {journal}
  {\bibinfo  {journal} {Entropy}\ }\textbf {\bibinfo {volume} {23}} (\bibinfo
  {year} {2021}{\natexlab{b}}),\ 10.3390/e23081073}\BibitemShut {NoStop}%
\bibitem [{\citenamefont {Aaronson}\ and\ \citenamefont
  {Gottesman}(2004)}]{aaronson2004ImprovedSimulationStabilizer}%
  \BibitemOpen
  \bibfield  {author} {\bibinfo {author} {\bibfnamefont {S.}~\bibnamefont
  {Aaronson}}\ and\ \bibinfo {author} {\bibfnamefont {D.}~\bibnamefont
  {Gottesman}},\ }\href {\doibase 10.1103/PhysRevA.70.052328} {\bibfield
  {journal} {\bibinfo  {journal} {Phys. Rev. A}\ }\textbf {\bibinfo {volume}
  {70}},\ \bibinfo {pages} {052328} (\bibinfo {year} {2004})}\BibitemShut
  {NoStop}%
\bibitem [{\citenamefont {Bravyi}\ and\ \citenamefont
  {Gosset}(2016)}]{bravyi2016ImprovedClassicalSimulation}%
  \BibitemOpen
  \bibfield  {author} {\bibinfo {author} {\bibfnamefont {S.}~\bibnamefont
  {Bravyi}}\ and\ \bibinfo {author} {\bibfnamefont {D.}~\bibnamefont
  {Gosset}},\ }\href {\doibase 10.1103/PhysRevLett.116.250501} {\bibfield
  {journal} {\bibinfo  {journal} {Phys. Rev. Lett,}\ }\textbf {\bibinfo
  {volume} {116}},\ \bibinfo {pages} {250501} (\bibinfo {year}
  {2016})}\BibitemShut {NoStop}%
\bibitem [{\citenamefont {Yoshida}\ and\ \citenamefont
  {Kitaev}(2017)}]{yoshida2017EfficientDecodingHaydenPreskill}%
  \BibitemOpen
  \bibfield  {author} {\bibinfo {author} {\bibfnamefont {B.}~\bibnamefont
  {Yoshida}}\ and\ \bibinfo {author} {\bibfnamefont {A.}~\bibnamefont
  {Kitaev}},\ }\href {\doibase 10.48550/arXiv.1710.03363} {\enquote {\bibinfo
  {title} {Efficient decoding for the {{Hayden-Preskill}} protocol},}\ }
  (\bibinfo {year} {2017}),\ \Eprint {http://arxiv.org/abs/1710.03363}
  {arXiv:1710.03363 [hep-th, physics:quant-ph]} \BibitemShut {NoStop}%
\bibitem [{Note2()}]{Note2}%
  \BibitemOpen
  \bibinfo {note} {We refer to ``query access'' as the ability to perform the
  unitary transformation $U_t$ followed by a measurement on a quantum register
  consisting of $n$ qubits}\BibitemShut {NoStop}%
\bibitem [{\citenamefont {Leone}\ \emph {et~al.}(2024)\citenamefont {Leone},
  \citenamefont {Oliviero}, \citenamefont {Lloyd},\ and\ \citenamefont
  {Hamma}}]{leone2022LearningEfficientDecoders}%
  \BibitemOpen
  \bibfield  {author} {\bibinfo {author} {\bibfnamefont {L.}~\bibnamefont
  {Leone}}, \bibinfo {author} {\bibfnamefont {S.~F.~E.}\ \bibnamefont
  {Oliviero}}, \bibinfo {author} {\bibfnamefont {S.}~\bibnamefont {Lloyd}}, \
  and\ \bibinfo {author} {\bibfnamefont {A.}~\bibnamefont {Hamma}},\ }\href
  {\doibase 10.1103/PhysRevA.109.022429} {\bibfield  {journal} {\bibinfo
  {journal} {Phys. Rev. A}\ }\textbf {\bibinfo {volume} {109}},\ \bibinfo
  {pages} {022429} (\bibinfo {year} {2024})}\BibitemShut {NoStop}%
\bibitem [{\citenamefont {Yoshida}\ and\ \citenamefont
  {Yao}(2019)}]{yoshida2019DisentanglingScramblingDecoherence}%
  \BibitemOpen
  \bibfield  {author} {\bibinfo {author} {\bibfnamefont {B.}~\bibnamefont
  {Yoshida}}\ and\ \bibinfo {author} {\bibfnamefont {N.~Y.}\ \bibnamefont
  {Yao}},\ }\href {\doibase 10.1103/PhysRevX.9.011006} {\bibfield  {journal}
  {\bibinfo  {journal} {Phys. Rev. X}\ }\textbf {\bibinfo {volume} {9}},\
  \bibinfo {pages} {011006} (\bibinfo {year} {2019})}\BibitemShut {NoStop}%
\bibitem [{Note3()}]{Note3}%
  \BibitemOpen
  \bibinfo {note} {The simplified settings of this Letter correspond to a
  special class of $t$-doped Clifford circuits. In particular, one can always
  write $U_t=C_0 W_{1}W_{2}\protect \cdots W_t C_{0}^{\prime }$, where
  $C_{0},C_{0}^{\prime }$ are Clifford, while $W_{j}$ are $1$-doped Clifford
  circuits. The setting explored in this Letter corresponds to the special case
  where $[W_{j},W_{k}]=0$ for $j,k\in {1,t/2}$, and $W_{j+t/2}=C_{j}^{\dagger
  }W_{j}C_{j}$ for $j\in {1,\protect \ldots ,t/2}$, such that
  $[W_{j+t/2},W_{j}]\protect \neq 0$, with ${C_{j}}$ being a set of commuting
  Clifford operators. In this manner, if $P$ is the Pauli generator not
  preserved by the action of $W_{j}$, then $C_{j}^{\dagger }PC_{j}$ is also not
  preserved.}\BibitemShut {Stop}%
\bibitem [{See()}]{SeeSupplementalMaterial}%
  \BibitemOpen
  \href@noop {} {\enquote {\bibinfo {title} {See supplemental material.
  including the following
  citations~\cite{yoshida2017EfficientDecodingHaydenPreskill,yoshida2019DisentanglingScramblingDecoherence,oliviero2021TransitionsEntanglementComplexity,leone2022RetrievingInformationBlack}},}\
  }\BibitemShut {NoStop}%
\bibitem [{Note4()}]{Note4}%
  \BibitemOpen
  \bibinfo {note} {Note that this probability is different from the one
  displayed in Eq.~\protect \eqref {successprobability}. The reason is that
  Čebyšëv inequality allows us to say that $\protect \mathcal
  {F}(V)=\protect \frac {1}{1+2^{2|A|-2|E|}}+O(\varepsilon )$ with probability
  $O(\varepsilon ^{-2}2^{-2|C|})$, while the more detailed analysis of~\cite
  {leone2022LearningEfficientDecoders} shows $\protect \mathcal {F}(V)\ge
  \protect \frac {1}{1+2^{2|A|-2|E|}}$ (see Eq.~\protect \eqref {mainF}) with
  probability given by Eq.~\protect \eqref {successprobability}.}\BibitemShut
  {Stop}%
\bibitem [{\citenamefont
  {Gottesman}(1998)}]{gottesman1998HeisenbergRepresentationQuantuma}%
  \BibitemOpen
  \bibfield  {author} {\bibinfo {author} {\bibfnamefont {D.}~\bibnamefont
  {Gottesman}},\ }in\ \href
  {http://citeseerx.ist.psu.edu/viewdoc/summary?doi=10.1.1.252.9446} {\emph
  {\bibinfo {booktitle} {International {{Conference}} on {{Group Theoretic
  Methods}} in {{Physics}}}}}\ (\bibinfo {year} {1998})\BibitemShut {NoStop}%
\bibitem [{\citenamefont {Harlow}(2023)}]{harlow2023blackholesquantumgravity}%
  \BibitemOpen
  \bibfield  {author} {\bibinfo {author} {\bibfnamefont {D.}~\bibnamefont
  {Harlow}},\ }\href {https://arxiv.org/abs/2304.10367} {\enquote {\bibinfo
  {title} {Black holes in quantum gravity},}\ } (\bibinfo {year} {2023}),\
  \Eprint {http://arxiv.org/abs/2304.10367} {2304.10367} \BibitemShut {NoStop}%
\bibitem [{\citenamefont {Yang}\ and\ \citenamefont
  {Engelhardt}(2023)}]{yang2023ComplexityLearningPseudo}%
  \BibitemOpen
  \bibfield  {author} {\bibinfo {author} {\bibfnamefont {L.}~\bibnamefont
  {Yang}}\ and\ \bibinfo {author} {\bibfnamefont {N.}~\bibnamefont
  {Engelhardt}},\ }\href {https://arxiv.org/abs/2302.11013} {\enquote {\bibinfo
  {title} {The complexity of learning (pseudo)random dynamics of black holes
  and other chaotic systems},}\ } (\bibinfo {year} {2023}),\ \Eprint
  {http://arxiv.org/abs/2302.11013} {2302.11013} \BibitemShut {NoStop}%
\bibitem [{\citenamefont
  {Page}(1993{\natexlab{b}})}]{page1993AverageEntropySubsystem}%
  \BibitemOpen
  \bibfield  {author} {\bibinfo {author} {\bibfnamefont {D.~N.}\ \bibnamefont
  {Page}},\ }\href {\doibase 10.1103/PhysRevLett.71.1291} {\bibfield  {journal}
  {\bibinfo  {journal} {Phys. Rev. Lett,}\ }\textbf {\bibinfo {volume} {71}},\
  \bibinfo {pages} {1291} (\bibinfo {year} {1993}{\natexlab{b}})}\BibitemShut
  {NoStop}%
\bibitem [{\citenamefont {Leone}\ \emph {et~al.}(2022)\citenamefont {Leone},
  \citenamefont {Oliviero}, \citenamefont {Piemontese}, \citenamefont {True},\
  and\ \citenamefont {Hamma}}]{leone2022RetrievingInformationBlack}%
  \BibitemOpen
  \bibfield  {author} {\bibinfo {author} {\bibfnamefont {L.}~\bibnamefont
  {Leone}}, \bibinfo {author} {\bibfnamefont {S.~F.~E.}\ \bibnamefont
  {Oliviero}}, \bibinfo {author} {\bibfnamefont {S.}~\bibnamefont
  {Piemontese}}, \bibinfo {author} {\bibfnamefont {S.}~\bibnamefont {True}}, \
  and\ \bibinfo {author} {\bibfnamefont {A.}~\bibnamefont {Hamma}},\ }\href
  {\doibase 10.1103/PhysRevA.106.062434} {\bibfield  {journal} {\bibinfo
  {journal} {Phys. Rev. A}\ }\textbf {\bibinfo {volume} {106}},\ \bibinfo
  {pages} {062434} (\bibinfo {year} {2022})}\BibitemShut {NoStop}%
\end{thebibliography}
\end{document}


\title{Unscrambling quantum information with Clifford decoders - Supplemental materials}

\author{Salvatore F.E. Oliviero}\email{s.oliviero001@umb.edu}
\affiliation{Physics Department,  University of Massachusetts Boston,  02125, USA}
\affiliation{Theoretical Division (T-4), Los Alamos National Laboratory, Los Alamos, New Mexico 87545, USA}
\affiliation{Center for Nonlinear Studies, Los Alamos National Laboratory, Los Alamos, New Mexico 87545, USA}

\author{Lorenzo Leone}\email{lorenzo.leone001@umb.edu}
\affiliation{Physics Department,  University of Massachusetts Boston,  02125, USA}
\affiliation{Theoretical Division (T-4), Los Alamos National Laboratory, Los Alamos, New Mexico 87545, USA}
\affiliation{Center for Nonlinear Studies, Los Alamos National Laboratory, Los Alamos, New Mexico 87545, USA}

\author{{Seth} {Lloyd}}\email{slloyd@mit.edu}
\affiliation{Department of Mechanical Engineering, Massachusetts Institute of Technology,  {Cambridge},  {MA}, {USA}}
\affiliation{Turing Inc., Brooklyn, NY, USA}

\author{Alioscia Hamma}\email{alioscia.hamma@unina.it}

\affiliation{Dipartimento di Fisica `Ettore Pancini', Universit\`a degli Studi di Napoli Federico II,
Via Cintia 80126,  Napoli, Italy}
\affiliation{INFN, Sezione di Napoli, Italy}

\maketitle

\setcounter{secnumdepth}{2}
\setcounter{equation}{0}
\setcounter{figure}{0}
\renewcommand{\thetable}{S\arabic{table}}
\renewcommand{\theequation}{S\arabic{equation}}
\renewcommand{\thefigure}{S\arabic{figure}}
\titleformat{\section}[hang]{\normalfont\bfseries}
{Supplemental Material \thesection:}{0.5em}{\centering}

\setcounter{equation}{0}
\setcounter{figure}{0}
\setcounter{table}{0}

\makeatletter
\renewcommand{\theequation}{S\arabic{equation}}
\renewcommand{\thefigure}{S\arabic{figure}}
\renewcommand{\bibnumfmt}[1]{[S#1]}
\newtheorem{thmS}{Theorem S\ignorespaces}

\newtheorem{claimS}{Claim S\ignorespaces}

We provide proofs and additional details supporting the claims in the main text.

\tableofcontents

\section{Decoupling through randomization: Proof of Eq.~$(11)$}
\begin{figure*}[th!]
    \centering
    \includegraphics[width=\textwidth]{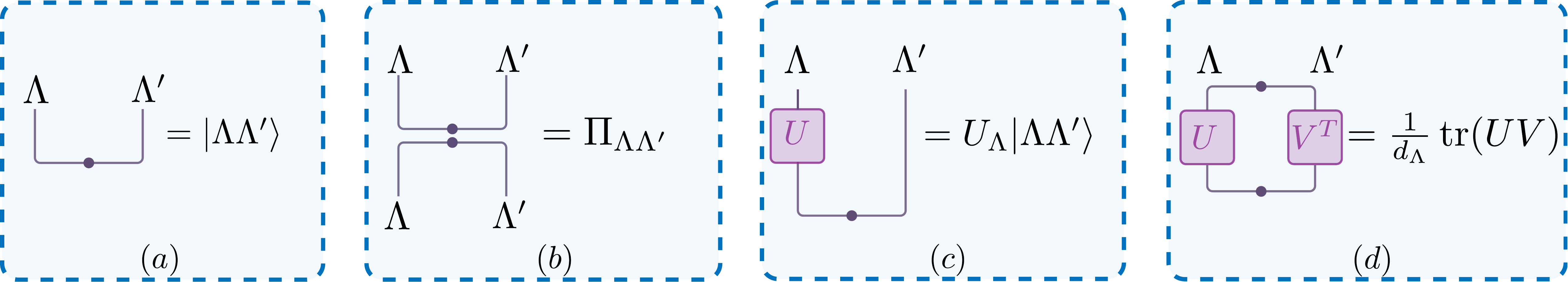}
    \caption{Diagrammatic picture of (a) an EPR pair $\ket{\Lambda\Lambda^\prime}$, (b) a projector $\Pi_{\Lambda\Lambda^\prime}\equiv \ket{\Lambda\Lambda^\prime}\bra{\Lambda\Lambda^\prime}$ (c) the action of a unitary operator $U$ on the EPR pair $\ket{\Lambda\Lambda^\prime}$ (d) the expectation value of the product of the operators $U$ and $V$ in the completely mixed state over the system $\Lambda$}
    \label{rules}
\end{figure*}

In this section, we prove that, with high probability, the randomizer $\mathcal{R}$, acting on the subspaces $F^{'}$ and $C^{'}$, returns the fidelity in Eq.~$(11)$ by inducing a decoupling effect. In the following, each state will be associated with a diagram. In Fig.~\ref{rules} we show a few introductive diagrams that will help in the comprehension of the next results.

Let us consider the output state $\ket{\psi_{V}}$
\begin{equation}
\label{appeq:psiout}   
\ket{\psi_V}=\frac{1}{\sqrt{\pi_{V}}}\figbox{.22}{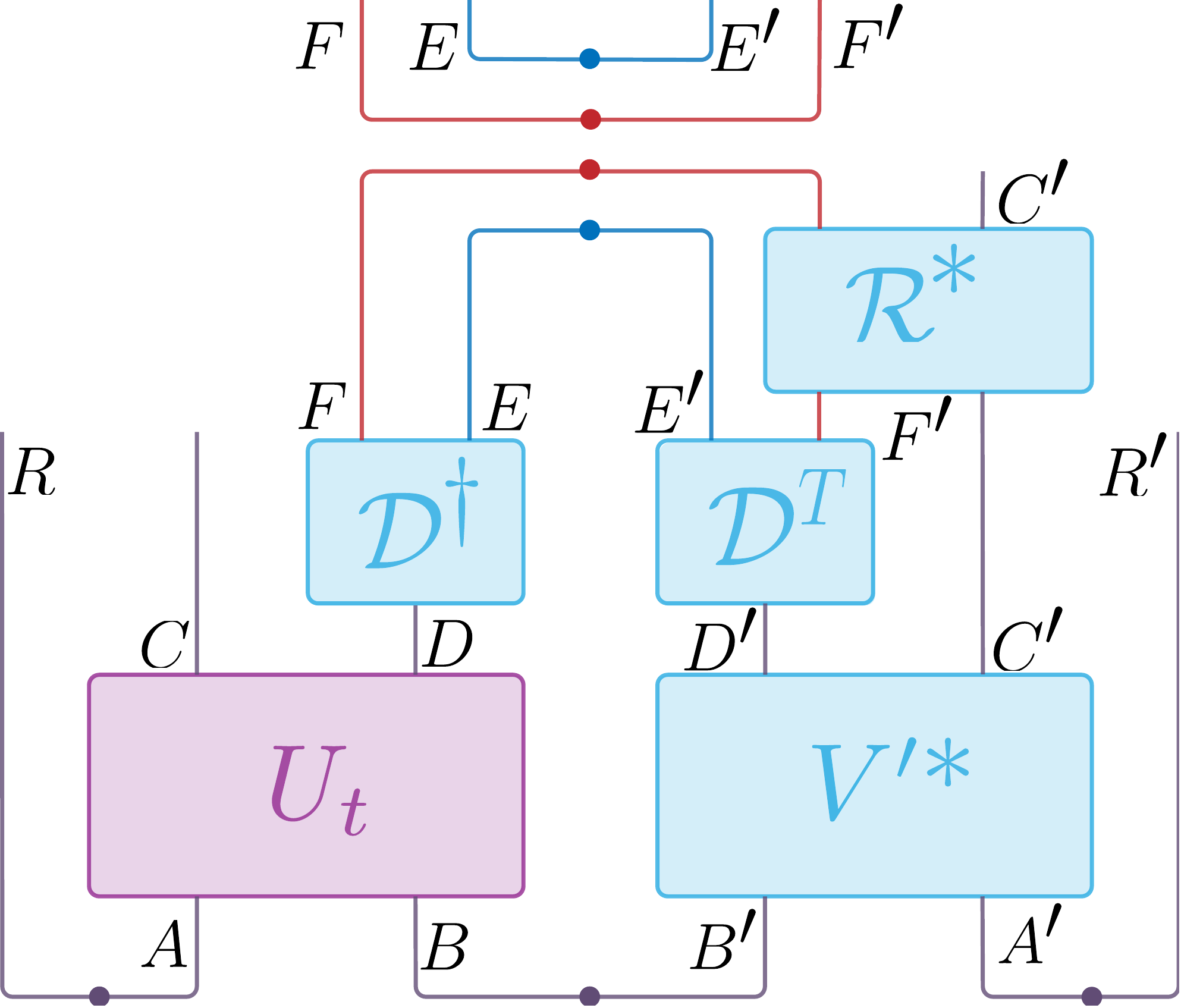}=\frac{1}{\sqrt{\pi_{V}}}\Pi_{DD^\prime}(\mathcal{R}^*_{C^\prime F^{\prime}}\mathcal{D}^{T}_{D^{\prime}}V^{\prime *}_{A^\prime B^{\prime}})\otimes (\mathcal{D}_D^\dag [U_t]_{AB})\ket{BB^{\prime}}\ket{AR}\ket{A^{\prime}R^{\prime}}
\end{equation}
where $P_{V}$ is the normalization coefficient of $\ket{\psi_V}$. To prove that the randomizer $\mathcal{R}$ is able to decouple the composite subsystem $EB^{\prime}$ we have to look at the behavior of the fidelity. To compute the fidelity of the output state $\psi_V$ with the input state, we have to project the output state $\ket{\psi_V}\bra{\psi_V}$ on the projector $\Pi_{RR^\prime}$:
\ba
\label{appeq:fid}
\mathcal{F}(V)=\tr(\Pi_{RR^\prime}\ket{\psi_V}\bra{\psi_V})=\frac{\mathcal{F}(V)\pi_{V}}{{\pi_{V}}}=\frac{\sum_{P_D}\tr(P_D(\mathcal{D}^\dag U_t))P_D(\mathcal{R}\mathcal{D}^\dag V^\prime))}{\sum_{P_DP_A}\tr(P_AP_D(\mathcal{D}^\dag U_t))P_AP_D(\mathcal{R}\mathcal{D}^\dag V^\prime))}
\ea
where we wrote it as the ratio between $\mathcal{F}(V)\pi_V$ and $\pi_V$, because has shown in~\cite{yoshida2017EfficientDecodingHaydenPreskill,yoshida2019DisentanglingScramblingDecoherence}, both possess a diagram that, thanks to the average over the Pauli group and Bell pairs, can be written in terms of OTOCs, returning the \emph{r.h.s}. A proof of Eq.~\eqref{appeq:fid} can be found in~\cite{leone2022RetrievingInformationBlack}. As stated in the incipit of the section our main interest is the effect of the randomizer in the learning protocol. Let us set up the following notation, given a subsystem $\Lambda$, $|\Lambda|$ will label the number of qubits of the subsystem $\Lambda$. A way to study the randomizer is by looking at its average behavior and its fluctuations. Define the following coefficients
\ba
N_1&:=&2^{2|A|} \pi_V \mathcal{F}(V) =\frac{1}{2^{2|D|+n}}\sum_{P_D}\tr(P_D(\mathcal{D}^\dag U_t))P_D(\mathcal{R}\mathcal{D}^\dag V^\prime))\,,\\
N_2&:=&2^{2|A|} \pi_V =\frac{1}{2^{2|D|+n}}\sum_{P_D,P_A}\tr(P_AP_D(\mathcal{D}^\dag U_t))P_AP_D(\mathcal{R}\mathcal{D}^\dag V^\prime))\,.
\ea

Then the fidelity $\mathcal{F}(V)$ can be rewritten as a function of $N_1$ and $N_2$ as $\mathcal{F}(V)=N_1/N_2$. Due to the correlations between $N_1$ and $N_2$, average over $\mathcal{F}(V)$ and its fluctuations we consider the following expansion for $\aver{\mathcal{F}(V)}_{\mathcal{R}}$
\be
\aver{\mathcal{F}(V)}_{\mathcal{R}}=\frac{\aver{N_1}}{\aver{N_2}}+\Theta\left(\frac{\aver{N_1N_2}-\aver{N_1}\aver{N_2}}{\aver{N_2}^2}\right)\,
\ee
and the following one for the fluctuations $\Delta_{\mathcal{R}}\mathcal{F}(V)\equiv\aver{\mathcal{F}^2(V)}-\aver{\mathcal{F}(V)}^2$ around the average 
\be
\Delta_{\mathcal{R}}\mathcal{F}(V)=\Theta\left[\frac{\Delta_{\mathcal{R}} N_1}{\aver{N_2}^2}+\frac{\Delta_{\mathcal{R}} N_2 \aver{N_1}^{2}}{\aver{N_2}^4}-2\frac{(\aver{N_1N_2}-\aver{N_1}\aver{N_2})N_1}{\aver{N_2}^3}+3 \left(\frac{\aver{N_1N_2}-\aver{N_1}\aver{N_2}}{\aver{N_2}^2}\right)^2\right]\,,
\ee
a complete proof of these expansions can be found in~\cite{oliviero2021TransitionsEntanglementComplexity}.
It is now necessary to compute each coefficient in $\aver{\mathcal{F}(V)}_{\mathcal{R}}$ and $\Delta_{\mathcal{R}}\mathcal{F}(V)$.
\subsection{Average of $N_1$ and $N_2$}
Let us compute the average of $N_1$. First, let us give a diagrammatic picture of this average
\ba
\aver{N_1}&=&\int \de \mathcal{R}\,\figbox{.2}{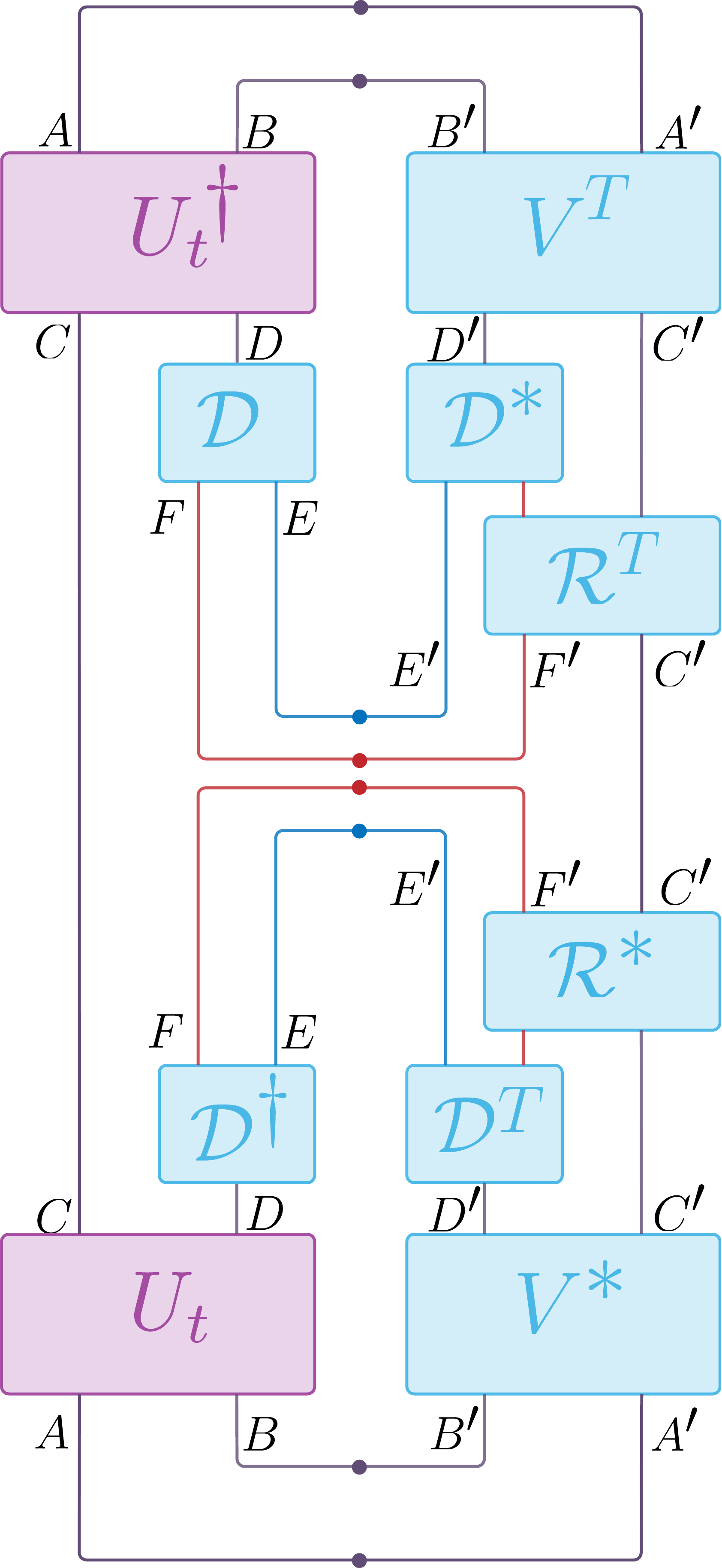}=\figbox{.2}{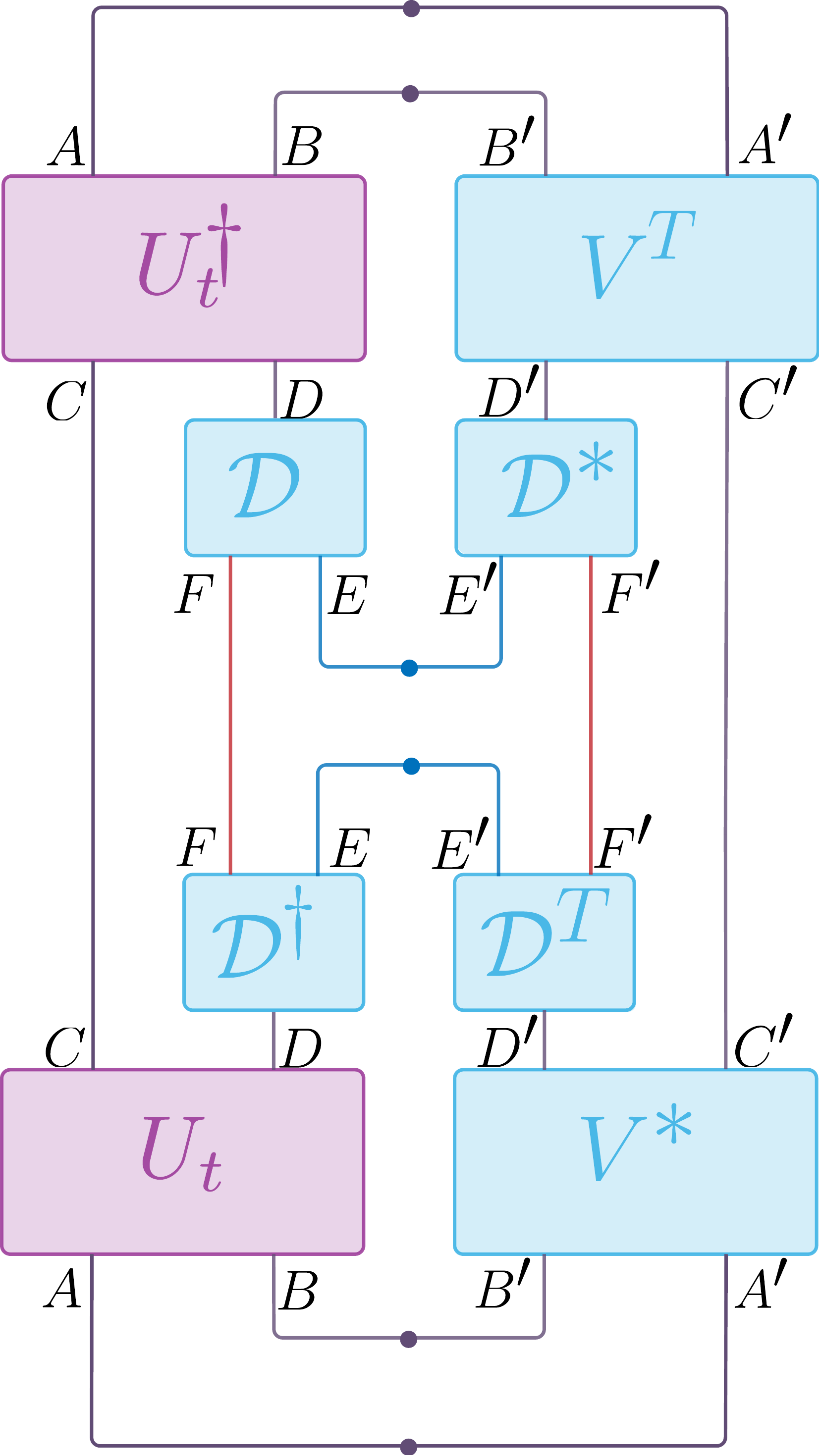}\\
&=&2^{-n}\aver{\tr\left(P_D(\mathcal{D}^\dag U_t)P_D(\mathcal{D}^{\dag}V)\right)}_{P_D}
\label{eq19}
\ea

where the averaged diagram is given by the trace of the product between $\Pi_{RR^\prime}$ with the unnormalized $\st{\psi_V}$ times a multiplicative factor, while the \emph{r.h.s} shows the effect of the average of $\mathcal{R}$ on the diagram. Let us expand Eq.~\eqref{eq19} the sum of a function depending on $P_D$ as $\sum_{P_D}f(P_D)=\sum_{P_E,P_F}f(P_E\otimes P_F)=\sum_{P_{E}}f(P_E\otimes \bbbone_{F})+\sum_{P_{E},P_F\neq\bbbone_{F}}f(P_E\otimes P_F)$, it means that the average on $N_1$ can be written as 
\ba
\aver{N_{1}}&=&2^{-2|D|-n}\sum_{P_E}\tr(P_{E}(\mathcal{D}^\dag U_t)P_{E}(\mathcal{D}^\dag V))+2^{-2|D|-n}\sum_{P_E,P_F\neq\bbbone_{F}}\tr\left[P_{E}\otimes P_F(\mathcal{D}^\dag U_t)\aver{P_{E}\otimes P_F(\mathcal{R}\mathcal{D}^\dag V)}_{\mathcal{R}}\right]\nonumber\\
&=&2^{-2|D|-n}\sum_{P_E}\tr(P_{E}(\mathcal{D}^\dag U_t)P_{E}(\mathcal{D}^\dag V))
\ea

Where we used the fact that $\aver{P_{F}}_{\mathcal{R}}=0$ for $P_F\neq\bbbone_{F}$. For $N_2$ diagrammatically we have
\ba
\int \de\mathcal{R}~~ N_2 &=&2^{2|A|-n}\int \de \mathcal{R}\,\figbox{.22}{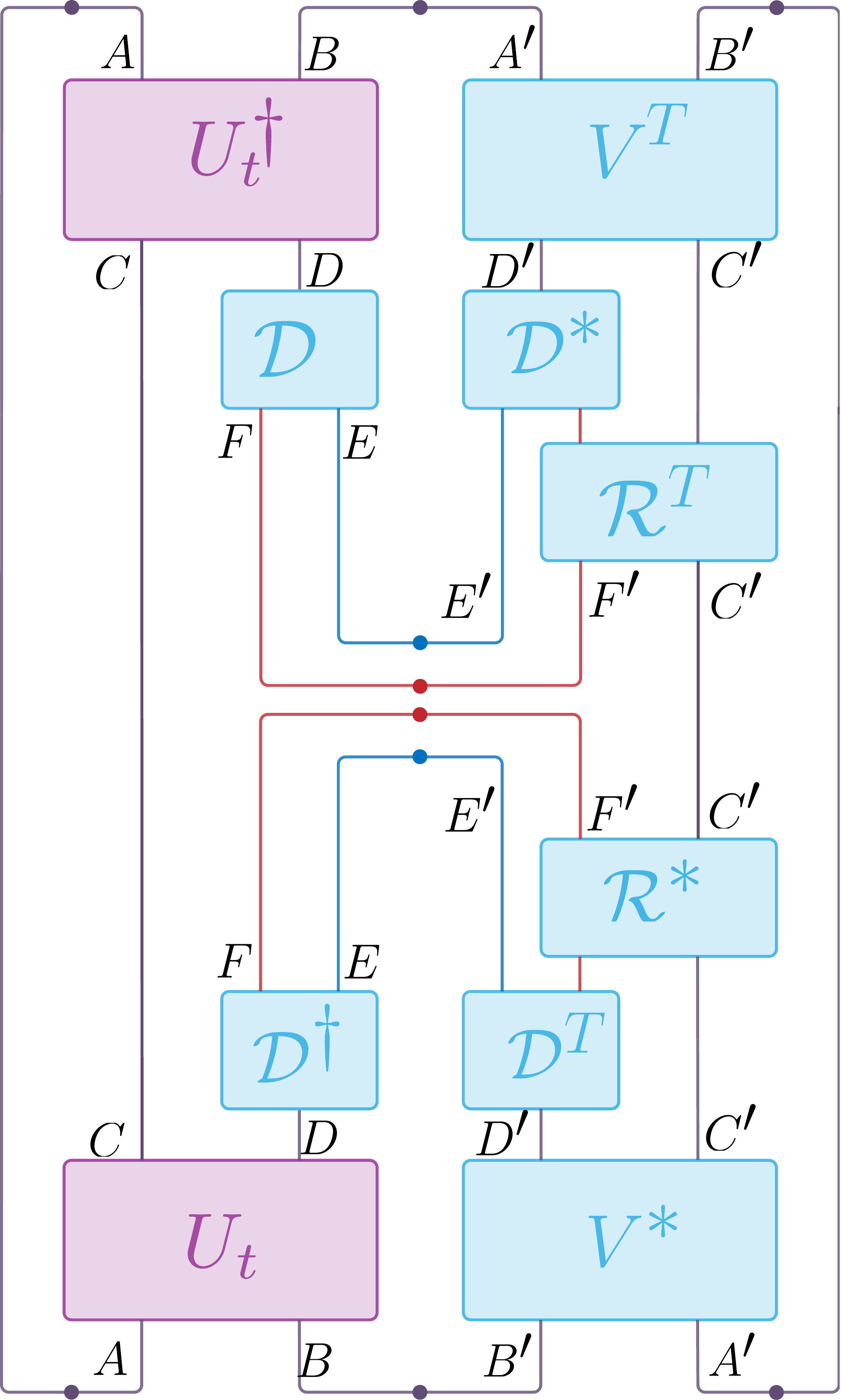}=2^{-n}\,\figbox{.22}{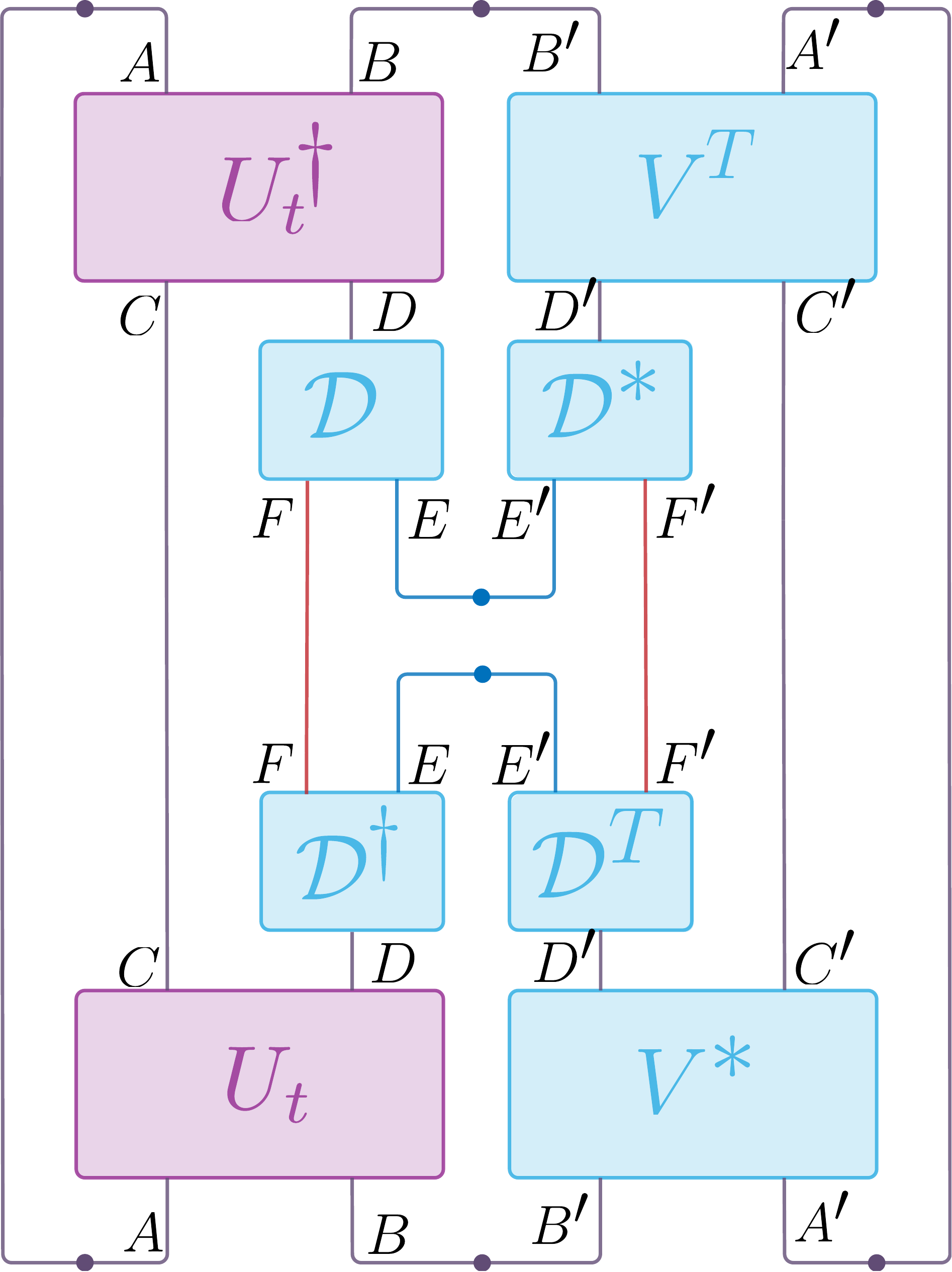}\\
&=&2^{2|A|-n}\aver{\tr\left(P_D(\mathcal{D}^\dag U_t)P_AP_D(\mathcal{D}^{\dag}V)P_A\right)}_{P_D,P_A}
\ea 

that corresponds to the normalization coefficient of $\ket{\psi_V}$ times a multiplicative factor. Repeating the calculations done for $\aver {N_1}$, we obtain
\ba
\aver{N_{2}}&=&2^{-2|D|-n}\sum_{P_E,P_A}\tr(P_AP_{E}(\mathcal{D}^\dag U_t)P_AP_{E}(\mathcal{D}^\dag V))
\ea

When we make the assumption that our unitary operator $U_t$ is a scrambling unitary and in the hypothesis of $(\mathcal{D}^\dag U_t)^{\dag}P_E \mathcal{D}^\dag U_t=(\mathcal{D}^\dag V)^{\dag}P_E \mathcal{D}^\dag V$ for all $P_E\in \mathbb{P}_E$, we obtain
\ba
\aver{N_1}&=&2^{-2|D|-n}\sum_{P_E}\tr(P_{E}(\mathcal{D}^\dag U_t)P_{E}(\mathcal{D}^\dag V))=2^{-2|F|}\\
\aver{N_2}&=&2^{-2|D|-n}\sum_{P_E,P_A}\tr(P_{E}(\mathcal{D}^\dag U_t)P_AP_{E}(\mathcal{D}^\dag U_t) P_A\simeq \frac{1}{2^{2|D|}}(2^{2|E|}+2^{2|A|}-1)
\ea 
\subsection{Correlation terms}
Computed the average of $N_1$ and $N_2$, we can then proceed to calculate higher-order moments. Let us look at the average of the product between $N_1$ and $N_2$
\ba
\aver{N_1N_2}&=&2^{-4|D|-2n}\sum_{P_E,Q_E,P_A}\tr(P_{E}(\tilde{U})P_{E}(\tilde{V}))\tr(P_AQ_{E}(\tilde{U})P_AQ_{E}(\tilde{V}))\\
&+&2^{-4|D|-2n}\!\!\!\!\!\!\!\!\!\sum_{P_E,Q_E,P_A,P_F\neq\bbbone_F}\!\!\!\!\!\!\!\!\!\!\!\tr(P_{E}\otimes P_F(\tilde{U})\aver{P_{E}\otimes P_F(\mathcal{R}\tilde{V})}_{\mathcal{R}})\tr\left[P_AQ_{E}(\tilde{U})P_AQ_{E}(\tilde{V})\right]\nonumber\\
&+&2^{-4|D|-2n}\!\!\!\!\!\!\!\!\!\sum_{P_E,Q_E,P_A,Q_F\neq\bbbone_F}\!\!\!\!\!\!\!\!\!\!\!\tr(P_{E}(\tilde{U})P_{E} (\tilde{V}))\tr\left[P_AQ_{E}\otimes Q_F(\tilde{U})P_A \aver{Q_{E}\otimes Q_F(\mathcal{R}\tilde{V})}_{\mathcal{R}}\right]\nonumber\\
&+&2^{-4|D|-2n}\!\!\!\!\!\!\!\!\!\!\!\!\sum_{P_E,Q_E,P_A,P_FQ_F\neq\bbbone_F}\!\!\!\!\!\!\!\!\!\!\!\!\!\!\!\!\!\!\tr\left[(P_{E}\otimes P_F(\tilde{U})\otimes P_AQ_{E}\otimes Q_F(\tilde{U})P_A )\aver{P_{E}\otimes P_F (\mathcal{R}\tilde{V})\otimes Q_{E}\otimes Q_F (\mathcal{R}\tilde{V})}_{\mathcal{R}}\right]\nonumber
\ea
where we defined $\tilde{U}=\mathcal{D}^\dag U_t$ and $\tilde{V}=\mathcal{D}^\dag V$.
Performing the average over $\mathcal{R}$ one can easily observe, that the first term is equivalent to the product of $\aver{N_1}$ and $\aver{N_2}$, while the next two terms are equal to zero because $\aver{P_F}_\mathcal{R}=0$ for $P_F\neq \bbbone$, and defining $\tilde{U}=\mathcal{D}^\dag U_t$ and $\tilde{V}=\mathcal{D}^\dag V$, we obtain
\ba\label{eqboh}
\aver{N_1N_2}&=&\aver{N_1}\aver{N_2}\\
&+&2^{-4|D|-2n}\!\!\!\!\!\!\!\sum_{P_E,Q_E,P_A,P_FQ_F\neq\bbbone_F}\!\!\!\!\!\!\!\tr\left[(\tilde{U}^{\dag}P_{E} P_F\tilde{U}\otimes P_A\tilde{U}^{\dag}Q_{E} Q_F\tilde{U}P_A )\aver{\tilde{V}^{\otimes 2}\mathcal{R}^{\dag\otimes 2}P_{E} P_F \otimes Q_{E} Q_F \mathcal{R}^{\otimes 2}\tilde{V}^{\otimes 2}}_{\mathcal{R}}\right]\nonumber
\ea

Let us take compute the average of $\mathcal{R}_{FC}^{\dag\otimes 2} P_F\otimes Q_F \mathcal{R}_{FC}^{\otimes 2}$. For any $P_F\neq \bbbone$ 
\be
\int d\mathcal{R}_{FC} \mathcal{R}_{FC}^{\dag\otimes 2} P_F\otimes Q_F \mathcal{R}_{FC}^{\otimes 2} =\delta_{QP}\left(\frac{2^{|F|+|C|} T_{FC}-\bbbone}{2^{2|F|+2|C|}-1}\right)
\ee
where $T_{FC}$ is the swap operator on the combined subsystem $CF$. We can then rewrite the second term of Eq.~\eqref{eqboh} as
\be
\frac{1}{2^{2|F|+2|C|}-1}\sum_{P_E,Q_E,P_A,P_FQ_F\neq\bbbone_F}\tr[(\tilde{U}^{\dag}P_{E} P_F\tilde{U}\otimes P_A\tilde{U}^{\dag}Q_{E} P_F\tilde{U}P_A )\tilde{V}^{\dag\otimes 2}(P_{E}\otimes Q_E) (2^{|F|+|C|}T_{FC}-\bbbone_{FC}^{\otimes 2})\tilde{V}^{\otimes 2}]
\ee

Now, let us consider the following hypothesis $U^{\dag}_t\mathcal{D}^{\dag}P_E \mathcal{D}^\dag U_t=V^{\dag}\mathcal{D}^{\dag}P_E \mathcal{D}^\dag V$. In this hypothesis the term related to $-\bbbone^{\otimes 2}_{CF}$ is null, consequently we obtain that

\ba
& &\frac{1}{2^{2|F|+2|C|}-2^{|E|}}\!\!\!\!\!\!\!\sum_{P_E,Q_E,P_A,P_FQ_F\neq\bbbone_F}\!\!\!\!\!\!\!\tr[(\tilde{U}^{\dag}P_{E} P_F\tilde{U}\otimes P_A\tilde{U}^{\dag}Q_{E} P_F\tilde{U}P_A )\tilde{V}^{\dag\otimes 2}(P_{E}\otimes Q_E) (2^{|F|+|C|}T_{FC}-\bbbone_{FC}^{\otimes 2})\tilde{V}^{\otimes 2}]\nonumber\\
&=&\frac{2^{|F|+|C|}}{2^{2|F|+2|C|}-2^{|E|}}\!\!\!\!\!\!\!\sum_{P_E,Q_E,P_A,P_FQ_F\neq\bbbone_F}\!\!\!\!\!\!\!\tr[(\tilde{U}^{\dag}P_{E} P_F\tilde{U}\otimes P_A\tilde{U}^{\dag}Q_{E} P_F\tilde{U}P_A )\tilde{V}^{\dag\otimes 2}(P_{E}\otimes Q_E) T_{FC}TT\tilde{V}^{\otimes 2}]\\
&=&\frac{2^{|F|+|C|}}{2^{2|F|+2|C|}-2^{|E|}}\!\!\!\!\!\!\!\sum_{P_E,Q_E,P_A,P_FQ_F\neq\bbbone_F}\!\!\!\!\!\!\!\tr[(\tilde{U}^{\dag}P_{E} P_F\tilde{U}\otimes P_A\tilde{U}^{\dag}Q_{E} P_F\tilde{U}P_A )\tilde{V}^{\dag\otimes 2}(P_{E}\otimes Q_E) T_{E}\tilde{V}^{\otimes 2}T]\nonumber
\ea
Where we used $TT_{CF}=T_E$ and $[T,\tilde{V}^{\otimes 2}]=0$; expanding $T_{E}=d_{E}^{-1}\sum_{K_E}K_{E}\otimes K_E$, we obtain
\ba
&=&\frac{2^{|C|+|F|}}{2^{|E|+2|C|+2|F|}-2^{|E|}}\sum_{P_E,Q_E,P_A,P_FQ_F\neq\bbbone_F}\tr[(\tilde{U}^{\dag}P_{E} P_F\tilde{U}\otimes P_A\tilde{U}^{\dag}Q_{E} P_F\tilde{U}P_A )\tilde{V}^{\dag\otimes 2}(P_{E}K_E\otimes Q_EK_E) \tilde{V}^{\otimes 2}T]\nonumber\\
&=&\frac{2^{|C|+|F|}}{2^{|E|+2|C|+2|F|}-2^{|E|}}\sum_{P_E,Q_E,P_A,P_FQ_F\neq\bbbone_F}\tr[ \tilde{U}^{\dag}P_{E} P_F\tilde{U}\tilde{V}^{\dag}P_{E}K_E\tilde{V}P_A\tilde{U}^{\dag}Q_{E} P_F\tilde{U}P_A\tilde{V}^{\dag}Q_{E}K_E\tilde{V}]
\ea
Since $\tilde{U}^{\dag}P_E \mathcal{D}\tilde{U}=\tilde{V}^{\dag}P_E \tilde{V}$ for all $P_E\in \mathbb{P}_E$, we can substitute the whole term with $U$:
\ba
&=&\frac{2^{|C|+|F|}}{2^{|E|+2|C|+2|F|}-2^{|E|}}\sum_{P_E,Q_E,P_A,P_FQ_F\neq\bbbone_F}\tr[ \tilde{U}^{\dag}P_{E} P_F\tilde{U}\tilde{U}^{\dag}P_{E}K_E\tilde{U}P_A\tilde{U}^{\dag}Q_{E} P_F\tilde{U}P_A\tilde{U}^{\dag}Q_{E}K_E\tilde{U}]\nonumber\\
&=&\frac{2^{3|E|+|C|+|F|}}{2^{2|C|+2|F|}-1}\sum_{Q_E,P_F\neq\bbbone, P_A}\tr[\tilde{U}^{\dag}Q_E P_F\tilde{U}P_A\tilde{U}^{\dag}Q_E P_F\tilde{U}P_A]\nonumber\\
\ea
Thus, in the hypothesis of $U$ being scrambling
\be
\aver{N_1N_2}\simeq \frac{2^{3|E|+|C|+|F|}}{2^{n+4|D|}(2^{2|C|+2|F|}-1)}(2^{2|D|}-2^{2|E|})+\aver{N_1}\aver{N_2}
\ee

We obtain
\be
\frac{\aver{N_1N_2}-\aver{N_1}\aver{N_2}}{\aver{N_2}^{2}}=\frac{2^{4|E|} \left(2^{2|F|}-1\right)}{\left(2^{2|A|}+2^{2|E|}-1\right)^2 \left(2^{2|F|+2|C|}-1\right)}=\mathcal{O}(2^{-2|C|})
\ee

Now, let us compute $\Delta_{\mathcal{R}} N_1\equiv \aver{N_1^2}-\aver{N_1}^2$, the calculations are similar to the one made for $\aver{N_1N_2}-\aver{N_1}\aver{N_2}$, we can repeat the calculation splitting the sum over the subsystems $E$ and $F$ one can easily show that
\ba
\Delta_{\mathcal{R}} N_1=\frac{1}{2^{2n+4|D|}}\sum\tr[(\tilde{U}^{\dag}P_{E} P_F\tilde{U}\otimes \tilde{U}^{\dag}Q_{E} Q_F\tilde{U} )\aver{\tilde{V}^{\otimes 2}\mathcal{R}^{\dag\otimes 2}P_{E} P_F \otimes Q_{E} Q_F \mathcal{R}^{\otimes 2}\tilde{V}^{\otimes 2}}_{\mathcal{R}}]\,.
\ea
Using the same techniques as before,$\Delta_{\mathcal{R}} N_1$ can be written as
\ba
\Delta_{\mathcal{R}} N_1&=&\frac{1}{2^{2n+4|D|}}\frac{2^{|C|+|F|}}{d_E(2^{2|F|+2|C|}-1)}\sum_{P_E,Q_E,Q_F,K_EP_F\neq \bbbone_F}\tr[ \tilde{U}^{\dag}P_{E} P_F\tilde{U}\tilde{U}^{\dag}P_{E}K_E\tilde{U}\tilde{U}^{\dag}Q_{E} P_F\tilde{U}\tilde{U}^{\dag}Q_{E}K_E\tilde{U}]\nonumber\\
&=&\frac{1}{2^{2n+4|D|}}\frac{2^{|C|+|F|}}{2^{|E|}(2^{2|F|+2|C|}-1)}\sum_{P_E,Q_E,Q_F,K_EP_F\neq \bbbone_F}\tr[ P_{E} P_FP_{E}K_EQ_{E} P_FQ_{E}K_E]\nonumber\\
&=&\frac{1}{2^{n+4|D|}}\frac{2^{|C|+|F|}}{2^{|E|}(2^{2|F|+2|C|}-1)}(2^{2|F|}-1)2^{6|E|}
\ea
A similar calculation can be carried for $\Delta_{\mathcal{R}} N_2$ returning
\be
\Delta_{\mathcal{R}} N_2=\frac{1}{2^{2n+4|D|}}\sum\tr[(P_A\tilde{U}^{\dag}P_{E} P_F\tilde{U}P_A\otimes Q_A\tilde{U}^{\dag}Q_{E} Q_F\tilde{U} Q_A)\aver{\tilde{V}^{\otimes 2}\mathcal{R}^{\dag\otimes 2}P_{E} P_F \otimes Q_{E} Q_F \mathcal{R}^{\otimes 2}\tilde{V}^{\otimes 2}}_{\mathcal{R}}]
\ee
Making the hypothesis of $U$ scrambling and $\tilde{V}^{\dag}\mathbb{P}_E\tilde{V}=\tilde{U}^{\dag}\mathbb{P}_E\tilde{U}$ we obtain
\ba
\Delta_{\mathcal{R}} N_2&=&\frac{1}{2^{2n+4|D|}}\frac{2^{|C|+|F|}\sum}{2^{|E|}(2^{2|C|+2|F|}-1)}\tr[ Q_A\tilde{U}^{\dag}P_{E} P_F\tilde{U}Q_A\tilde{V}^{\dag}P_{E}K_E\tilde{V}P_A\tilde{U}^{\dag}Q_{E} P_F\tilde{U}P_A\tilde{V}^{\dag}Q_{E}K_E\tilde{V}]\nonumber\\
&\simeq&\frac{1}{2^{n+4|D|}}\frac{2^{3|E|}2^{|C|+|F|}}{2^{2|C|}2^{2|F|}-1}(2^{2|D|}-2^{2|E|})
\ea
Note that $\aver{N_1}=\frac{1}{2^{2|F|}}$ for $\tilde{V}^{\dag}\mathbb{P}_E\tilde{V}=\tilde{U}^{\dag}\mathbb{P}_E\tilde{U}$. We obtain
\be
\Delta_{\mathcal{R}}\mathcal{F}=\frac{\left(2^{2|A|}-1\right)^2 2^{4|E|} \left(2^{2|F|}-1\right)}{\left(2^{2|A|}+2^{2|E|}-1\right)^4 \left(2^{2|C|+2|F|}-1\right)}=\mathcal{O}\left(2^{-2|C|}\right)
\ee
Note that $|A|=\mathcal{O}(1)$ and $|E|=\mathcal{O}(1)$ because, thanks to the decoupling theorem, one has $|E|=|A|+\log\epsilon^{-1/2}$. 
\subsubsection{Chebyshev inequality}
The last step is to prove that the probability to be away from the average is small, we make use of the Chebyshev inequality for the fidelity $\mathcal{F}_V(U)$ and we obtain
\be
\operatorname{Pr}\left[|\mathcal{F}_V(U)-\aver{\mathcal{F}_V(U)}_{\mathcal{R}}|\ge\epsilon\right]\le \frac{\Delta_{\mathcal{R}} \mathcal{F}}{\epsilon^2}= \mathcal{O}\left(\frac{2^{-2|C|}}{\epsilon^2}\right)
\ee